\documentclass[10pt,aps,physrev,ppreprint,superscriptaddress]{revtex4-2}
\usepackage[utf8]{inputenc}
\usepackage{physics}
\usepackage{amsmath,latexsym,amsfonts,amssymb,amsthm}
\usepackage{enumitem}
\usepackage{graphicx}
\usepackage{bbold}
\usepackage[margin=1in]{geometry}
\usepackage{comment}

\graphicspath{{Images/}}

\begin{document}

\title{The Geometry of Contraction-Induced Flows}
\author{Aaron Winn}
\email{winna@sas.upenn.edu}
\affiliation{Department of Physics and Astronomy, University of Pennsylvania, Philadelphia, Pennsylvania 19104, USA}
\author{Eleni Katifori}
\affiliation{Department of Physics and Astronomy, University of Pennsylvania, Philadelphia, Pennsylvania 19104, USA}
\affiliation{Center for Computational Biology, Flatiron Institute, New York, New York 10010, USA}

\date{\today}

\begin{abstract}
    Peristalsis is the driving mechanism behind a broad array of biological and engineered flows. In peristaltic pumping, a wave-like contraction of the tube wall produces local changes in volume which induce flow. Net flow arises due to geometric nonlinearities in the momentum equation, which must be properly captured to compute the flow accurately. While most previous models focus on radius-imposed peristalsis, they often neglect longitudinal length changes — a natural consequence of radial contraction in elastic materials. In this paper, to capture a more accurate picture of peristaltic pumping, we calculate the flow in an elastic vessel undergoing contractions in the transverse and longitudinal directions simultaneously, keeping the geometric nonlinearities arising in the strain. A careful analysis requires us to study our fluid using the Lagrangian coordinates of the elastic tube. We perform analytic calculations of the flow characteristics by studying the fluid inside a fixed boundary with time-dependent metric. This mathematical manipulation works even for large-amplitude contractions, as we confirm by comparing our analytical results to COMSOL simulations. We demonstrate that transverse and longitudinal contractions induce instantaneous flows at the same order in wall strain, but in opposite directions. We investigate the influence of the wall’s Poisson ratio on the flow profile. Incompressible walls suppress flow by minimizing local volume changes, whereas auxetic walls enhance flow. For radius-imposed peristaltic waves, wall incompressibility reduces both reflux and particle trapping. In contrast, length-imposed waves typically generate backflow, although trapping can still occur at large amplitudes for some Poisson ratios. These results yield a more complete description of peristalsis in elastic media and offer a framework for studying contraction-induced flows more broadly.
\end{abstract}

\maketitle

\section{Introduction}

Peristalsis, defined as fluid flow driven by the cyclic contraction of a tube wall, is ubiquitous in biological systems. Peristaltic waves drive fluid transport in the digestive system \cite{Nicosia2001LocalUltrasonography, Pal2002TheTransport,BrasseurEtAl2007}, ureter \cite{Carew_Pedley_1997, Kalayeh_Xie_Brian_Fowlkes_Sack_Schultz_2023}, lymphatic system \cite{Moore2018LymphaticFlows, Wolf2021FluidValves, Winn2024OperatingValves}, and perivascular spaces of the brain \cite{Carr2021PeristalticTubes, Trevino2025LowSpace}. Much theoretical work has been done to characterize peristaltic flows, typically in the context of radius-imposed peristalsis. In the models of radius-imposed peristalsis, the transverse motion of the tube wall is imposed and propagates as a wave, while the longitudinal wall motion is fixed to zero. Burns and Parkes presented an early theoretical treatment of radius-imposed peristalsis and were able to solve for the stream function for arbitrary wavelength deformations \cite{Burns_Parkes_1967}, but only at small amplitude. Transverse oscillations in a 2D planar geometry were the focus of the paper, though corrections for a tube with longitudinal displacements derived by enforcing tube inextensibility were briefly calculated. Incorporating inextensibility acknowledges that the deforming boundary is not an arbitrary mathematical function, but that it should actually obey physical material constraints. This approach was also taken to study the related problem of a swimming flagellated micro-organism \cite{Taylor_1951}. However, the majority of papers released since these pioneering works have ignored the mechanical properties of the wall, and Burnes and Parkes even noted that the difference between the problem with purely transverse oscillations and that where inextensibility is applied can be ignored \cite{Burns_Parkes_1967}. In this work we will see that, particularly in cases of large-amplitude peristalsis beyond where the perturbative approach of Burnes and Parkes is valid, longitudinal wall motion can have substantial consequences on peristaltic flow.

Soon after the work of Burnes and Parkes, Shapiro, et al. further analyzed radius-imposed peristalsis, relaxing the assumption of small-amplitude deformation, but only at long wavelengths in an infinitely long tube under the lubrication approximation \cite{Shapiro_Jaffrin_Weinberg_1969}. This led to an exact expression for the average flow given an arbitrary pressure drop per wavelength. The trick to solving for peristaltic flow is to recognize that if the boundary propagates as a wave, then the flow is steady in the co-moving frame. This property not only allows for analytic calculation of the flow, but also simplifies the calculation of particle trajectories. Shapiro, et al. highlight two interesting phenomena: reflux, where particles close to the wall travel against the wave direction, and trapping, where at large amplitude a bolus forms in which particles are trapped and move on average at the wave speed. A summary of early work on radius-imposed peristalsis including finite Reynolds number effects is given in \cite{Jaffrin_Shapiro_1971}. It is not too difficult to generalize the calculations to a finite tube, though the flow in that case is inherently non-steady \cite{Li1993Non-SteadyTubes}. Later numerical studies relaxed some or all assumptions regarding amplitude, wavelength, and Reynolds number, for the case of a 2D geometry \cite{Pozrikidis_1987} \cite{Takabatake_Ayukawa_1982} and a cylindrical geometry \cite{Takabatake_Ayukawa_Mori_1988}. All of these studies highlight the importance of keeping the nonlinear radius scaling in the conductance since the trapping phenomenon can only be observed at large amplitude where the geometric nonlinearities are important. However, all of these studies also neglect the mechanics of the tube wall and longitudinal wall motion.

Contrasted with radius-imposed peristalsis, in force-imposed peristalsis, a wave-like force is imposed on the exterior of a tube wall, and the boundary displacements are solved for alongside the fluid variables \cite{Tang1993NumericalBoundaries}. This is one way in which elastic tube properties can be incorporated into models of peristaltic pumping. Takagi and Balmforth show that radius-imposed peristalsis can be considered a special case of force-imposed peristalsis, where radial forces on the exterior of the tube balance with fluid force to displace the tube wall \cite{Takagi_Balmforth_2011}. Their paper highlights the important role that vessel stiffness plays in governing peristaltic flows. The tube radius will only deform in phase with the applied force when the vessel is sufficiently stiff. However, their work neglects longitudinal wall motion and shear forces. An even more complex class of peristalsis models separately treat passive elastic forces and active muscle contraction \cite{Carew_Pedley_1997, KouEtAl2015_ActiveMusculoMechanical}, though these models are limited to computational studies.  

To incorporate the effects of longitudinal wall motion, one needs to consider the coupling of the components of the strain tensor by the mechanics of the elastic wall. For a cylindrical tube modeled as a 2D axisymmetric membrane, there are two in-plane components of the strain tensor. Relaxing the symmetry constraint or allowing for finite tube thickness only further complicates the matter by introducing even more strains. These various coupled strains together determine the contraction-induced flow. In particular, we will see that longitudinal wall motion contributes to fluid flow at the same order in strain as transverse wall motion, despite being ignored in most cases. This longitudinal wall motion is well documented in the esophagus \cite{Nicosia2001LocalUltrasonography, Pal2002TheTransport,BrasseurEtAl2007,KouEtAl2015,KouEtAl2015_ActiveMusculoMechanical} and has also been described in the ureter \cite{Carew_Pedley_1997,Kalayeh_Xie_Brian_Fowlkes_Sack_Schultz_2023} and aorta \cite{AghilinejadEtAl2023}. Perhaps the clearest demonstration of coupled transverse and longitudinal peristaltic waves in biology can be seen in the locomotion of earthworms where longitudinally stretched regions correspond to regions of smaller diameter and longitudinally squished regions correspond to regions of larger diameter \cite{Quillin1999KINEMATICTERRESTRIS}. This same coupling occurs in the esophagus, which is believed to be incompressible: the muscle cross-sectional area and length are inversely proportional so that the the volume remains constant \cite{Nicosia2001LocalUltrasonography}. This simple coupling between strain components and its generalization to compressible walls has consequences on the induced fluid flow, which will be the focus of this paper. But, to understand this coupling, one needs to make some additional assumptions on the mechanics involved. For both the esophagus and the earthworm, one associates longitudinal contractions with longitudinal muscle activation and radial contractions with circumferential muscle. Both the active and passive wall properties are likely relevant for determining the full wall mechanics, but we will make some simple assumptions on the tube forces to arrive at an appropriate coupling between radial and longitudinal contractions. In particular, we will assume that the vessels are rigid enough or fluid forces are small enough that the fluid-to-solid coupling is negligible. Thus, our work generalizes the theory of radius-imposed peristalsis to include longitudinal wall motion, and should not be thought of as a generalization of force-imposed peristalsis, even though we will consider forces when determining how the strains are coupled together.

A few attempts at incorporating longitudinal strain into peristalsis calculations have been made, revealing some unique features of flows induced by longitudinal wall motion. Elbaz and Gat consider force-imposed peristalsis including external shear, but only consider linearized equations for the fluid and solid \cite{Elbaz_Gat_2014}. They demonstrate the role of the Poisson ratio, notably showing that shear forces do not induce flow in an incompressible tube. Trevino, et al. study a fully coupled fluid-structure interaction problem modeling the role of transverse and longitudinal wall motion in pumping cerebrospinal fluid near an elastic half space representing the brain \cite{Trevino2025LowSpace}. They find that longitudinal wall motion produces a net flow against the wave direction. Kalayeh, et al. focus their study on longitudinal effects in peristalsis by considering an Eulerian velocity boundary condition $\vec{V}^{\text{wall}}(\vec{x},t)$ which has both transverse and longitudinal components \cite{Kalayeh_Xie_Brian_Fowlkes_Sack_Schultz_2023}. They find that longitudinal wall motion can suppress reflux, when studied as a correction to radius-imposed peristalsis. This is the most complete theoretical work focusing on the role of longitudinal wall motion in peristaltic pumping, but they neglect the fact that the longitudinal and transverse wall motion should be coupled together, and instead impose the two components of $\vec{V}^{\text{wall}}(\vec{x},t)$ independently. Our theory reproduces some key results in these papers, but by considering a Lagrangian description of the wall, we are able to more carefully study the problem of contraction-induced flow with coupled longitudinal and transverse wall motion. 

The generalization of radius-imposed peristalsis to the case with longitudinal wall motion introduces new challenges in applying boundary conditions and analyzing flow. Unlike the case when only radial contractions are considered, if the tube is allowed to stretch longitudinally, the Eulerian coordinate $x$ specifying the axial position along the tube is no longer equal to the Lagrangian coordinate $X$ of a material point on the tube. Existing works studying the effect of longitudinal wall motion equate the fluid velocity to an imposed Eulerian wall velocity $\vec{V}^{\text{wall}}(\vec{x},t)$. This Eulerian approach is insufficient in capturing large-amplitude longitudinal effects and fluid-structure interaction. Instead, we will use the wall material (Lagrangian) coordinates $\vec{X}$ to parameterize the deformation of the solid tube: At time $t$, a Lagrangian coordinate $\vec{X}$ is located at position $\vec{x}_s(\vec{X},t)$. We will extend the Lagrangian coordinates of the solid into the fluid domain to conveniently describe the fluid using the Lagrangian coordinates of the solid. This greatly simplifies boundary conditions, but introduces a time-dependent metric which can be considered responsible for driving flows. Computationally, this method can be considered a special case of the Arbitrary Lagrangian-Eulerian Method (ALE), though its usefulness in allowing for analytical treatment of large-deformation fluid-structure interaction problems is so far unrecognized. We believe this method will be relevant for studying a broad class of contraction-induced flows analytically. Working with the Lagrangian coordinates of the solid possesses four advantages over a purely Eulerian approach, as outlined in the following paragraphs.

First, when evaluating the transport of contraction-driven flows, it is often useful to consider the volumetric flow rate, typically defined as the fluid flux through a fixed Eulerian surface. However, this will typically not be the relevant quantity. For example, when a theorist steps outside their office, about 5 liters of blood pass through the doorway in a half of a second, but a blood flow of 10 liters per second is a drastic overestimate of the flow relevant for delivering blood to tissues. The physiologically relevant flow through a particular blood vessel should be measured with respect to the walls of that vessel. A Lagrangian description of the wall motion makes this most explicit. 

Second, the Lagrangian description makes it easier to consider periodically deforming boundaries. Expressing the no-slip condition in Lagrangian coordinates is straightforward when the material deformation $\vec{x}_s(\vec{X},t)$ is known, but properly applying Eulerian boundary conditions is more subtle. To demonstrate this, consider a material boundary which returns to itself after some period $T$. One may hope that, at a fixed Eulerian position, having a wall velocity which $T$-averages to zero is sufficient to enforce this. However, one must evaluate the velocity along the path of a material particle. Thus, one is forced to explicitly consider the Lagrangian path of particles on the wall just to come up with a reasonable Eulerian velocity boundary condition. In practice, one would have to carefully tune a parameter in the function $\vec{V}^{\text{wall}}(\vec{x},t)$ to ensure periodicity is satisfied, as is done in \cite{Kalayeh_Xie_Brian_Fowlkes_Sack_Schultz_2023}, or else apply boundary conditions perturbatively \cite{Trevino2025LowSpace}. If instead, a Lagrangian description is used from the outset, then any $T$-periodic material deformation $\vec{x}_s(\vec{X},t)$ is appropriate. In general, applying the no-slip condition at a fluid-solid interface where the solid obeys a Lagrangian equation of motion is easiest using Lagrangian coordinates. 

Third, a material description of the boundary makes it easier to incorporate elastic effects. An Eulerian velocity boundary condition suggests that $\vec{V}^{\text{wall}}(\vec{x},t)$ can take any form desired, but elasticity theory constrains this. For most materials (like the earthworm), radial expansion will cause longitudinal shortening while radial contraction will cause longitudinal lengthening. This couples together the different components of $\vec{V}^{\text{wall}}$ such that $\partial_x V^{x,\text{wall}} \sim - V^{r,\text{wall}}/r$. The precise statement, particularly in cases of compliant vessels or large amplitude deformations, requires a careful treatment of the elastic boundary using the Lagrangian finite strain theory. 

Fourth, the Lagrangian description can easily be incorporated into networks. Fluid flow networks are inherently Lagrangian. One identifies pressures and flows at a node or edge, not at a particular coordinate in space. Thus, building network models out of Lagrangian theories, as we will do in this work, is natural. Networks of peristaltically contracting tubes have been used to model coordination of flows in the slime mold \cite{Alim2013RandomIndividual}, selective pumping in the insect tracheal network \cite{Aboelkassem2013SelectiveTransport}, and may be relevant for explaining fluid flows generated by swimming motion in the jellyfish gastrovascular system \cite{Southward1955ObservationsL.}. 

In this paper, we will study large-amplitude peristalsis in an axisymmetric tube with coupled circumferential and longitudinal strains by employing the Lagrangian coordinates of the material wall. In section two, our geometric approach for studying contraction-induced flows will be introduced. We derive the governing equations and calculate general solutions. In section three, we apply the model to a tube with spatially uniform strain. We argue that the radial and longitudinal displacements are coupled together through the Poisson ratio of the tube, and we study some basic properties of the fluid enclosed by this deforming tube. In section four, we employ this coupling between radial and longitudinal displacements to analyze peristaltically driven flows in tubes at different Poisson ratio. In section five, we briefly demonstrate how to apply these results to networks of deforming vessels. In section six, we give a summary of key results, and discuss generalizations and applications of the model. We elaborate on details of COMSOL simulations and flow calculations in the appendix.

\section{Fluid flow through a deforming elastic membrane}

\subsection{Basic Equations for the Fluid and Solid}

\begin{figure}
    \centering
    \includegraphics[width=.85\linewidth]{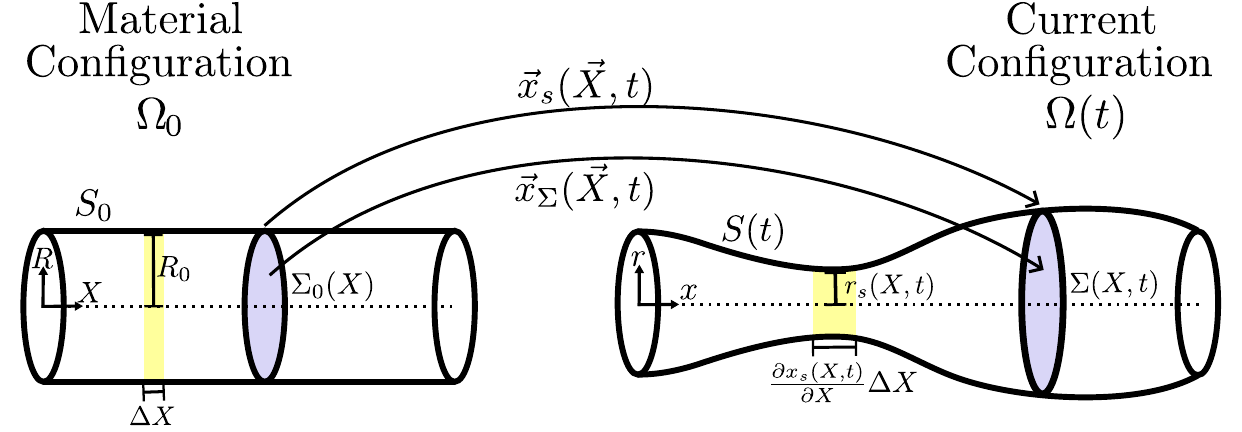}
    \caption{A fluid-filled cylindrical tube undergoing an arbitrary axisymmetric deformation. The material configuration $\Omega_0$ is described using the coordinates $\vec{X}(R,\Phi,X) = X\vec{e}_X + R\vec{e}_R(\Phi)$. When restricted to the solid boundary, these are the Lagrangian coordinates of the solid membrane $S_0$. The current configuration $\Omega(t)$ is described using the Eulerian coordinates $\vec{x}(r,\phi,x) = x \vec{e}_x + r \vec{e}_r(\phi)$. The surface $S_0$ is mapped to $S(t)$ via the map $\vec{x}_s$. This map is extended into the tube's interior via the map $\vec{x}_\Sigma$ which specifies how a surface in the fluid region deforms from $\Sigma_0(X)$ to $\Sigma(X,t)$. In this example, the left side of the tube is contracted radially and elongated longitudinally as illustrated by the yellow slice.}
    \label{fig:CoordinateSystems}
\end{figure}

Consider a cylindrical, fluid-filled tube, as shown in figure \ref{fig:CoordinateSystems}. At time $t$, the fluid occupies the volume $\Omega(t)$ and is bounded by an elastic membrane $S(t)$ and two open discs $\Sigma(X_1,t)$ and $\Sigma(X_2,t)$. The fluid in the tube is assumed to have zero Reynolds number and small radius $R_0$ when compared to its axial length scale (for a finite tube, the tube length $L_0$ or, for a peristaltic wave, the wavelength $\lambda$). Under these assumptions, commonly referred to as the lubrication approximation \cite{Tavakol2017ExtendedGeometry}, the continuity and momentum equations in $\Omega(t)$ read:
\begin{equation}
    \frac{\partial V^x}{\partial x}+ \frac{1}{r}\frac{\partial (rV^r)}{\partial r} = 0, \label{eq:Eulerian_incompressibility}
\end{equation}
\begin{align}
    \frac{\partial P}{\partial x} &= \frac{\mu}{r}\frac{\partial}{\partial r}\left(r \frac{\partial V^x}{\partial r}\right), \label{eq:dPdx}\\
    \frac{\partial P}{\partial r} &= 0 \label{eq:dPdr},
\end{align}
where $\vec{V}$ is the Eulerian fluid velocity field, $P$ is the pressure, and $\mu$ is the viscosity. 

Forgetting about the fluid for a moment, the displacement of the elastic membrane can be described using a Lagrangian map from a reference configuration $S_0$ to the current configuration $S(t)$. At rest, the tube has uniform radius $R_0$. A material (Lagrangian) coordinate in the reference configuration $\vec{X} \in S_0$ is mapped to an Eulerian coordinate $\vec{x} \in S(t)$ under the map $\vec{x} = \vec{x}_s(\vec{X},t)$. If the map is assumed to be axisymmetric,
\begin{equation}
    \vec{X}(\Phi,X,t) = X\vec{e}_X + R_0 \vec{e}_R (\Phi) \in  S_0,
\end{equation}
\begin{equation}
    \vec{x}_s(\Phi, X, t) = x_s(X,t)\vec{e}_X + r_s(X,t) \vec{e}_R(\Phi) \in  S(t),
\end{equation}
where $X$ is the position along the axial direction, and $\Phi$ is the angle. An example deformation is shown in figure \ref{fig:CoordinateSystems}. The displacement field is
\begin{equation}
    \vec{u}(\vec{X},t) = \vec{x}_s(\vec{X},t) - \vec{X} ,
\end{equation}
from which one can calculate the nonzero components of the in-plane strain tensor:
\begin{align}
    \epsilon_{\Phi \Phi} &=\hspace{.1cm} \frac{u_r}{ R_0}  + \frac{1}{2} \left(\frac{u_r}{ R_0}\right)^2, \label{eq:epsilon_PHIPHI}\\
    \epsilon_{XX}&= \frac{\partial u_x}{\partial X} + \frac{1}{2}\left( \frac{\partial u_x}{\partial X} \right)^2 + \frac{1}{2}\left( \frac{\partial u_r}{\partial X} \right)^2. \label{eq:epsilon_XX}
\end{align}
The tube may undergo large deformations, so the nonlinear finite strain is used, though $\frac{\partial u_r}{\partial X}$ is negligible by the lubrication approximation. In the membrane theory, the in-plane thickness-averaged stress (stress resultant divided by thickness) is related to the in-plane finite strain by Hooke's law:
\begin{align}
    \sigma_{\Phi \Phi} &= \frac{E}{1-\nu^2} (\epsilon_{\Phi \Phi} + \nu \epsilon_{XX}) , \label{eq:sigma_PHIPHI}\\
    \sigma_{XX} &= \frac{E}{1-\nu^2} (\epsilon_{XX} + \nu \epsilon_{\Phi \Phi}) \label{eq:sigma_XX}.
\end{align}
The Young's modulus $E$ will not play a role in the limits taken in this paper. The Poisson ratio $\nu$ will be the key parameter dictating how $\epsilon_{\Phi \Phi}$ and $\epsilon_{XX}$ are coupled. In particular, we will consider cases where one of the two stresses is zero. For $\nu > 0$, an increase in $\epsilon_{\Phi \Phi}$ causes a decrease in $\epsilon_{XX}$ and vice versa. For $\nu = 0$, there is no coupling between the strains. For $\nu < 0$, an increase in $\epsilon_{\Phi \Phi}$ causes an increase in $\epsilon_{XX}$ and vice versa. When $\nu = 1/2$, the tube is incompressible, meaning the bulk modulus diverges, and the local volume remains constant. Earlier works on peristalsis considered inextensible (length-preserving) 1D boundaries \cite{Taylor_1951, Burns_Parkes_1967, Walker2010ShapePumping}. Incompressibility (volume-preserving) is the equivalent property in the cylindrical geometry. Note that for $\nu \neq 0$, the membrane theory predicts a change in thickness $\epsilon_{RR}$, but we will only need to consider the in-plane strain for the analysis in this paper.

At large amplitudes many biological materials exhibit non-Hookean responses \cite{Fung1993}. In this paper, only geometric nonlinearities will be considered, while the stress-strain relationship is assumed linear. 

Assuming the stresses exerted by the fluid on the solid are small, we do not care to know the internal stresses of the solid, but the displacements directly couple to the fluid via the no-slip condition: 
\begin{equation}
    \vec{V}(\vec{x}_s(\vec{X},t),t) = \frac{\partial \vec{x}_s(\vec{X},t)}{\partial t}. \label{eq:no_slip_Lagrangian}
\end{equation}
In a purely Eulerian framework, the no-slip condition is written 
\begin{equation}
    \vec{V}(\vec{x},t) = \vec{V}^{\text{wall}}(\vec{x},t), \label{eq:no_slip_Eulerian}
\end{equation}
where $\vec{V}^{\text{wall}}(\vec{x},t)$ is a function representing wall motion, either imposed, or coupled to solid dynamics. As discussed in the introduction, the quantity  $\vec{V}^{\text{wall}}(\vec{x},t)$ cannot easily be expressed in terms of wall material properties. Once we have established the necessary machinery, we will avoid using boundary condition \eqref{eq:no_slip_Eulerian}.


In this section, it will be demonstrated how the Lagrangian coordinates of the solid can be used to unambiguously define the fluid flow. Then, by extending the material coordinates into the fluid domain, the fluid can be analyzed in the solid's rest frame. 

\subsection{A One-Dimensional Material Description of Contraction-Induced Flows}

As hinted in the introduction, an appropriate definition of the flow should subtract the motion of the boundary from the fluid motion. Specifically, flow should be measured with respect to a surface at fixed material coordinate $X$. To be precise, let $\Sigma_0(X)$ denote a surface of constant $X$ in $\Omega_0$, and let 
\begin{equation}
    \Sigma(X,t) \equiv \big{\{ }\vec{x} \in \Omega(t) \big{|} x = x_s(X,t), \hspace{.1cm} X \in \Sigma_0(X) \big{\} }
\end{equation}
denote the image of the surface $\Sigma_0(X)$ under the map $\vec{x}_s$, as shown in figure \ref{fig:CoordinateSystems}. Then, the flow through the surface $\Sigma(X,t)$ is
\begin{equation}
    q(X,t) \equiv \int_{\Sigma(X,t)} \left[\vec{V}(\vec{x},t) - \vec{V}^{\text{wall}}(\vec{x},t)\right]\cdot \hat{n} \hspace{.1cm}d\Sigma, \label{eq:flow_preliminary_definition}
\end{equation}
where $\hat{n}$ is the unit vector normal to the surface, oriented in the direction along our tube. For us, $\hat{n} = \vec{e}_X$, though one could even apply this formula to tubes undergoing bending or global rotation, in which case $\hat{n}$ will change with time. A more precise definition of $q(X,t)$ will be given in the next subsection, but to gain intuition for why this is an appropriate definition of the volumetric flow rate, integrate \eqref{eq:Eulerian_incompressibility} over the volume $\Omega(t)$ and apply the divergence theorem:
\begin{align}
    0 &= \int_{\Omega(t) }\vec{\nabla} \cdot \vec{V}(\vec{x},t) d\Omega \nonumber \\
    &= \int_{\partial \Omega(t)}  \left[\vec{V}(\vec{x},t) - \vec{V}^{\text{wall}}(\vec{x},t)\right]\cdot \hat{n} \hspace{.1cm} d^2x +  \int_{\partial \Omega(t)}  \vec{V}^{\text{wall}}(\vec{x},t)\cdot \hat{n} \hspace{.1cm} d^2x \nonumber \\
    &= \int_{\Sigma(X_2,t)}  \left[\vec{V}(\vec{x},t) - \vec{V}^{\text{wall}}(\vec{x},t)\right]\cdot \vec{e}_X d\Sigma - \int_{\Sigma(X_1,t)}  \left[\vec{V}(\vec{x},t) - \vec{V}^{\text{wall}}(\vec{x},t)\right]\cdot \vec{e}_X d\Sigma +  \frac{\partial \text{Vol}(\Omega(t))}{\partial t}  \nonumber \\
    &= q(X_2,t) - q(X_1,t)  + \frac{\partial \text{Vol}(\Omega(t))}{\partial t} \label{eq:dVoldt}.
\end{align}
In the third line, we used the no-slip boundary condition \eqref{eq:no_slip_Eulerian} to eliminate the integral of $\vec{V}(\vec{x},t) - \vec{V}^{\text{wall}}(\vec{x},t)$ over $S(t)$, and we noticed that the second integral in the second line is just the rate of change of volume of the enclosed region. Here, $X_1$ and $X_2$ are the material coordinates of the left and right endpoints of the tube. In the fourth line, we applied our definition of the flow \eqref{eq:flow_preliminary_definition}. As expected by physical intuition, any change in the volume of $\Omega(t)$ induces flow through the boundaries $\Sigma(X_1,t)$ and $\Sigma(X_2,t)$. 

We can also define a pressure in terms of the material coordinates of the solid. Equation \eqref{eq:dPdr} suggests that $P$ is only a function of $x$, so define 
\begin{equation}
    p(X,t) \equiv P(x_s(X,t),t)
\end{equation}
to be a pressure which is only a function of $X$. 

Taking the limit $(X_2-X_1) \rightarrow 0$ in equation \eqref{eq:dVoldt} and integrating the momentum equations \eqref{eq:dPdx} and \eqref{eq:dPdr} gives a one-dimensional system of equations purely in terms of the material coordinate $X$:
\begin{align}
    \frac{\partial q}{\partial X} + \frac{\partial }{\partial t} \left( \pi r_s^2 \frac{\partial x_s}{\partial X} \right)&= 0 \label{eq:continuity_integrated}\\
    \frac{\partial p}{\partial X} + \left(\frac{8\mu }{\pi r_s^4} \frac{\partial x_s}{\partial X}\right) q &= 0. \label{eq:momentum_integrated}
\end{align}
These two equations are a one-dimensional description of the fluid in terms of the material coordinate $X$. The first equation is an integrated form of the continuity equation. The second equation represents force balance, where a pressure gradient is balanced by viscous shear stress. The system is closed by applying boundary conditions on the pressure or flow. We will consider pressure boundary conditions. If the pressures at the left and right endpoint are known, and letting 
\begin{equation}
    \Delta p(t) \equiv p(X_2,t) - p(X_1,t),
\end{equation}
the flow at all material coordinates satisfies
\begin{equation}
    q(X,t) = \frac{-\Delta p(t)}{\int_{X_1}^{X_2} \frac{8\mu }{\pi r_s(X')^4} \frac{\partial x_s(X')}{\partial X} dX'} - \frac{\int_{X_1}^{X_2} \frac{8\mu }{\pi r_s(X')^4} \frac{\partial x_s(X')}{\partial X} \left(\int_{X'}^X \frac{\partial}{\partial t}\left(\pi r_s(X'')^2 \frac{\partial x_s(X'')}{\partial X} \right) dX'' \right)dX'}{\int_{X_1}^{X_2} \frac{8\mu}{\pi r_s(X')^4} \frac{\partial x_s(X')}{\partial X}  dX'}. \label{eq:q(X,t)_general}
\end{equation}
A derivation is provided in the appendix. This result is valid even for a tube undergoing large longitudinal strain or rigid body motions. Setting $\frac{\partial x_s}{\partial X}  = 1$ recovers the known result for contractions in a finite tube undergoing purely radial contractions \cite{Li1993Non-SteadyTubes}. 
Later sections will consider how particular constraints on the tube lead to particular forms of $q(X,t)$.

\subsection{A Three-Dimensional Material Description of Contraction-Induced Flows}

Integrating over a surface $\Sigma(X,t)$ moving with the tube produced a set of 1D equations for the pressure and flow. Taking this idea one step further, we wish to come up with a coordinate system describing points not only on the tube boundary, but also on the interior of the tube bounded by $\partial \Omega_0$. This is necessary in order to understand the fluid velocity and stress fields inside the tube. The domain of the undeformed, axisymmetric tube and its interior will be referred to as the material configuration and will be denoted $\Omega_0$. It is parameterized by the coordinates: 
\begin{equation}
    \vec{X}(R,\Phi, X, t) = X\vec{e}_X + R \vec{e}_R(\Phi) \in \Omega_0.
\end{equation}
As the boundary deforms, $\vec{X}(R_0,\Phi,X,t)$ is mapped to $\vec{x}_s(\Phi,X,t)$. How should coordinates in the interior of a fluid-filled tube deform? From our definition of $\Sigma(X,t)$, it is natural to define a surface $\Sigma_0(X)$ to be a surface of constant $X$ in $\Omega_0$. Specifying how points on $\Sigma_0(X)$ get mapped to points on $\Sigma(X,t)$ will associate with each material point $\vec{X} \in \Omega_0$ a point $\vec{x} \in \Omega(t)$. Denote this map $\vec{x}_\Sigma (\vec{X},t)$. The only requirement for this map is that it agrees with $\vec{x}_s(\vec{X},t)$ at $R=R_0$ and respects the axisymmetric assumption. The simplest choice is a map which is linear in $R$:
\begin{equation}
    \vec{x}_\Sigma (R,\Phi, X, t) = x_s(X,t) \vec{e}_x + r_s(X,t)\frac{R}{R_0} \vec{e}_r(\Phi). \label{eq:xSigma}
\end{equation}
The various geometric quantities are summarized in figure \ref{fig:CoordinateSystems}.  Note that the flow is defined as an integral over a surface and does not depend on the choice of $\vec{x}_\Sigma$. 

Now that we have a set of material coordinates to describe the fluid domain, we can define a velocity field in $\Omega_0$. Letting $\vec{X}_p(t)$ denote the $\vec{X}$ coordinates of particle $p$ at time $t$, the material velocity $\vec{v}$ is defined as
\begin{equation}
    \vec{v}(\vec{X}_p(t),t) \equiv \frac{d\vec{X}_p(t)}{dt}.
\end{equation}
It is linearly related to the Eulerian velocity $\vec{V}$ by a coordinate transformation:
\begin{equation}
    V^i(\vec{x}_\Sigma(\vec{X},t),t) = \sum_J \frac{\partial x_\Sigma^i(\vec{X},t)}{\partial X^J}v^J(\vec{X},t) + \frac{\partial x_\Sigma^i(\vec{X},t) }{\partial t} . \label{eq:Euler_V}
\end{equation}
In cylindrical coordinates, this reads
\begin{align}
    V^x(\vec{x}_\Sigma(\vec{X},t),t)
    &= \frac{\partial x_s(\vec{X},t)}{\partial X}v^X(\vec{X},t) + \frac{\partial x_s(\vec{X},t) }{\partial t}\\
    V^r(\vec{x}_\Sigma(\vec{X},t),t) &= \frac{R}{R_0}\left[\frac{\partial r_s(\vec{X},t)}{\partial X}v^X(\vec{X},t) + \frac{ r_s(\vec{X},t)}{R}v^R(\vec{X},t) + \frac{\partial r_s(\vec{X},t) }{\partial t}\right] .
\end{align}
Because boundary points in the material configuration are fixed, the no-slip boundary condition is simply
\begin{equation}
    \vec{v}(\vec{X},t) = 0 , \hspace{1cm} \vec{X} \in  S_0.
\end{equation}
Equation \eqref{eq:xSigma} induces a metric on $\Omega_0$ given by 
\begin{equation}
    g_{IJ} \equiv \frac{\partial \vec{x}_\Sigma}{\partial X^I} \cdot \frac{\partial \vec{x}_\Sigma}{\partial X^J},
\end{equation}
whose nonzero components for our cylindrical coordinate system are
\begin{equation}
    g_{RR} = \left( \frac{r_s}{R_0}\right)^2, \hspace{.5cm} g_{\Phi \Phi} = \left( \frac{r_s}{R_0}\right)^2 R^2, \hspace{.5cm} g_{XX} = \left( \frac{\partial x_s}{\partial X}\right)^2 + \left( \frac{R}{R_0} \frac{\partial r_s}{\partial X}\right)^2, \hspace{.5cm} g_{RX} = \frac{r_s}{R_0} \frac{R}{R_0} \frac{\partial r_s}{\partial X}
\end{equation}
and determinant is 
\begin{equation}
    \sqrt{|g|} = R \frac{\pi r_s^2}{\pi R_0^2} \frac{\partial x_s}{\partial X}.
\end{equation}
The factor of $R$ comes from the usual volume scaling in cylindrical coordinates, while the remaining factors describe the effective local growth in volume in the material configuration. Although the boundary $\partial \Omega_0$ is unchanging, the metric accounts for the changing geometry of the material configuration as summarized in the modified continuity equation:
\begin{equation}
    \frac{1}{\sqrt{|g|}}\frac{\partial \sqrt{|g|}}{\partial t} + \vec{\nabla} \cdot \vec{v} =0 \label{eq:continuity_material_clean}
\end{equation}
\begin{equation}
    \implies \frac{1}{\pi r_s^2 \frac{\partial x_s}{\partial X}} \frac{\partial (\pi r_s^2 \frac{\partial x_s}{\partial X})}{\partial t} + \frac{1}{R}\frac{\partial}{\partial R} (Rv^R) + \frac{1}{(\pi r_s^2 \frac{\partial x_s}{\partial X})}\frac{\partial}{\partial X} (\pi r_s^2 \frac{\partial x_s}{\partial X} v^X) = 0. \label{eq:continuity_material}
\end{equation}
This equation encompasses the primary difference between the fluid in $\Omega_0$ and that in $\Omega(t)$. In $\Omega(t)$, the fluid has zero divergence and is driven by the boundary. In $\Omega_0$, the fluid has a divergence: the change in metric acts as a source for the velocity field. 

The flow can now be defined purely in terms of material quantities as 
\begin{equation}
    q(X,t) \equiv  \int_0^{2\pi } \int_{0}^{R_0} v^X \sqrt{|g|} \hspace{.1cm}dR \hspace{.1cm} d\Phi, \label{eq:q_definition}
\end{equation}
which is equivalent to the previous definition \eqref{eq:flow_preliminary_definition}. An integral of \eqref{eq:continuity_material} over a surface of constant $X$ easily recovers the one-dimensional continuity equation \eqref{eq:continuity_integrated}. The full momentum equation can be treated in material coordinates using the techniques developed for studying fluid flows in curved space \cite{Aris1989}, though we will only need the specific case relevant for narrow, long tubes which is most easily derived by performing a coordinate transformation on equations \eqref{eq:dPdx} and \eqref{eq:dPdr}: 
\begin{align}
    \left(\frac{\partial x_s}{\partial X} \right)^{-2}\frac{\partial p}{\partial X} &= \mu
    \left(\frac{R_0}{r_s} \right)^2 \frac{1}{R} \frac{\partial}{\partial R} \left( R \frac{\partial v_X}{\partial R} \right)\\
    \frac{\partial p}{\partial R} &= 0.
\end{align}
Integrating this expression recovers \eqref{eq:momentum_integrated}. This three-dimensional formulation also allows us to solve for the velocity profiles generated by contracting channels. The velocity components satisfy
\begin{align}
    v^X(R,X,t) &= \frac{2q(X,t)}{\pi r_s(X,t)^2 \frac{\partial x_s(X,t)}{\partial X}}\left( 1- \frac{R^2}{R_0^2}\right) \label{eq:v^X_general}\\
    v^R(R,X,t) &= \frac{1}{\pi r_s(X,t)^2 \frac{\partial x_s(X,t)}{\partial X}} \frac{\partial}{\partial t} \left(\pi r_s(X,t)^2 \frac{\partial x_s(X,t)}{\partial X} \right) \frac{R}{2}\left(1 - \frac{R^2}{R_0^2} \right). \label{eq:v^R_general}
\end{align}
The magnitude of $v^X$ is maximal at $R=0$ while the magnitude of $v^R$ is maximal at $3^{-1/2}R_0 = 0.58 R_0$. Both components vanish at $R=R_0$, as required by the boundary conditions. Note that, so far, we have made no assumptions on the form of the boundary deformation, so equations \eqref{eq:v^X_general}, \eqref{eq:v^R_general}, and \eqref{eq:q(X,t)_general} can be applied to any boundary deformation. 

\begin{figure}
    \centering
    \includegraphics[width=.95\linewidth]{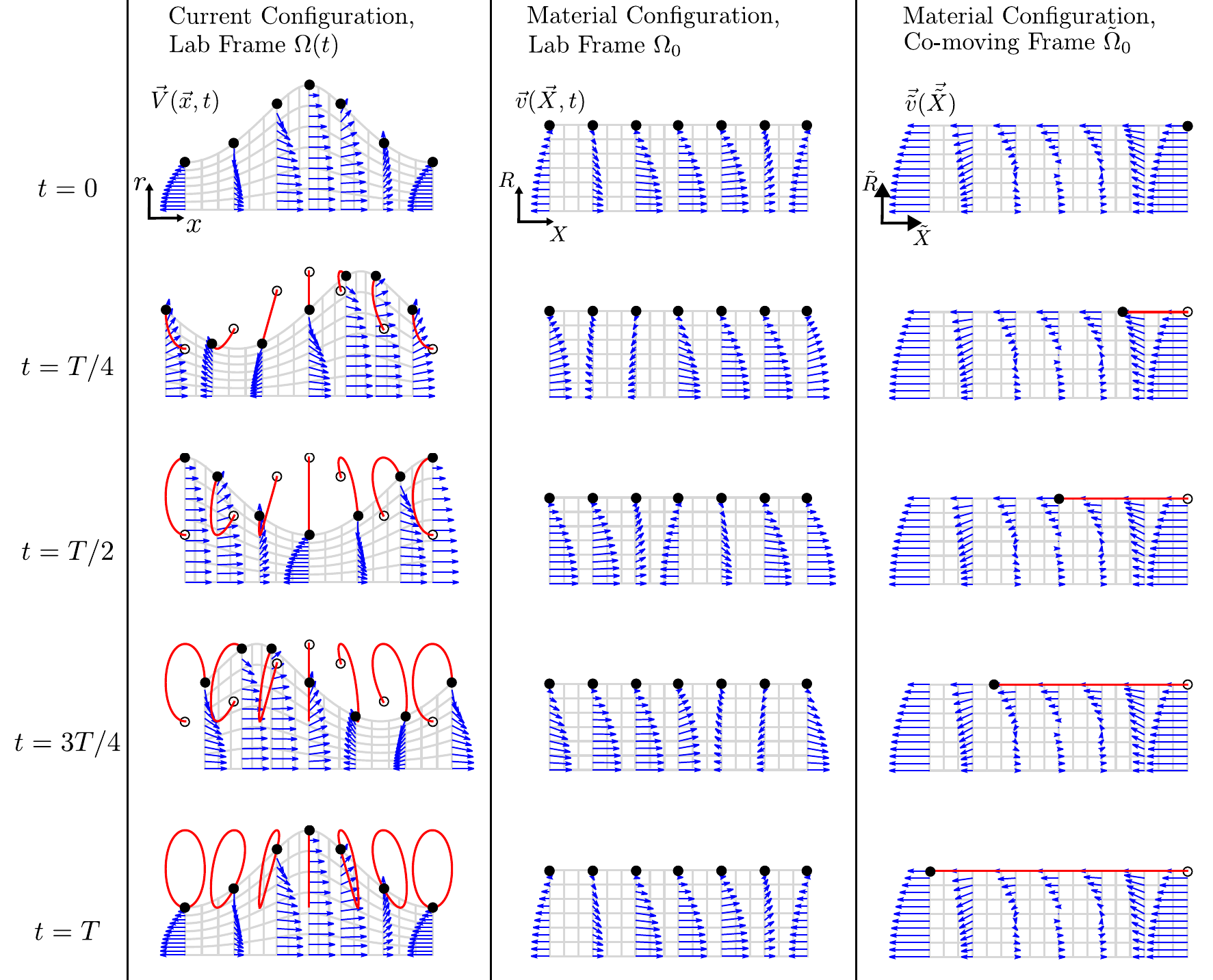}
    \caption{A tube is contracting peristaltically with period $T$. The velocity field is displayed in three different coordinate systems at five different time points: $t=0$, $t=T/4$, $t=T/2$, $t=3T/4$, and $t=T$. The trajectories of points on the wall are shown in red. Particles begin their trajectories at $t=0$ at the locations marked by hollow circles, and the current locations are marked by filled-in circles. In the left column, the Eulerian velocity field $\vec{V}(\vec{x},t)$ is plotted and points on the boundary are trace out loops. This represents the trajectories an observer would actually see. In the center column, the material velocity field $\vec{v}(\vec{X},t)$ is plotted, and the boundary points are stationary. The grid of lines of constant $X$ and $R$ in the material configuration deform to form the distorted grid in the current configuration. In the right column, the material velocity in the co-moving frame $\vec{\tilde{v}}(\vec{\tilde{X}})$ is plotted. The co-moving frame travels at velocity $c \vec{e}_X$, so particles on the boundary travel at velocity $-c \vec{e}_X$. The velocity field is independent of time in the co-moving frame. One could also consider Eulerian coordinates in the co-moving frame $\tilde{\Omega}(t)$, but this is not used in the paper. The wave satisfies equations \eqref{eq:dxsdX_longitudinal} and \eqref{eq:ur_longitudinal} with $\epsilon = 0.45$, $\nu=0.5$, and $\Delta \bar{p}_\lambda = 0$.}
    \label{fig:threeCoordinateSystems}
\end{figure}

An example velocity field is shown in figure \ref{fig:threeCoordinateSystems}. Notice the difference between the two quantities $\vec{V} \in \Omega (t)$ and $\vec{v}\in \Omega_0$. While $\vec{V}$ describes the physical velocity of an incompressible fluid put into motion by a deforming boundary, $\vec{v}$ describes a fictional velocity induced by a time-dependent metric in an undeforming domain. A simple linear coordinate transformation \eqref{eq:Euler_V} relates these two quantities. While $v^X$ has a parabolic profile proportional to the pressure gradient (middle column of figure \ref{fig:threeCoordinateSystems}), $V^x$ is the sum of a parabolic profile and a flat profile given by the wall axial velocity (left column of figure \ref{fig:threeCoordinateSystems}). It is $v^X$ which is relevant for calculating the flow $q$. When the boundary in $\Omega(t)$ or the metric in $\Omega_0$ takes the form of a peristaltic wave, a constant velocity field is observed in a frame moving at the wave speed. The co-moving frame (using material coordinates with fixed boundary) will be denoted $\tilde{\Omega}_0$, and the Galilean transformed velocity field which is independent of time will be denoted $\vec{\tilde{v}}(\vec{\tilde{X}})$. See the right column of figure \ref{fig:threeCoordinateSystems}. This frame is only defined for peristaltic waves, so we will return to this in section IV. 

\section{Uniform contractions in an elastic tube}
\subsection{General Properties of Uniform Contractions}
As a first application, consider a uniformly contracting tube:
\begin{align}
    \frac{\partial x_s(X,t)}{\partial X} &= \frac{l_s(t)}{L_0}\\
    r_s(X,t) &= r_s(t)
\end{align}
where $l_s(t)$ is the current length of the tube. The strain is constant in space. Uniform contractions are particularly convenient in that $\vec{x}_s(\vec{X},t)$ is invertible. For example, if $x_s(0,t) = 0$, then $ X(x,t) = (x/l_s(t)) L_0$.

At an instant in time when $\frac{d}{dt}\sqrt{|g|} \sim \frac{d}{dt}(\pi r_s^2 l_s)$ vanishes, the $\vec{X}$ divergence of $\vec{v}$ must also vanish. When $\frac{d}{dt}\sqrt{|g|} > 0$, the local volume is everywhere expanding, and thus the divergence must be negative. The intuition is that if the local space for the fluid to occupy is growing, then the fluid will rush in from outside to fill the space. Assuming the tube remains at rest at $X=0$, the instantaneous flow in equation \eqref{eq:q(X,t)_general} simplifies to
\begin{equation}
    q(X,t) = - \frac{\partial}{\partial t} \left(\pi r_s(t)^2 l_s(t) \right) \frac{X}{L_0}-\frac{\pi r_s(t)^4}{8\mu l_s(t)}  \Delta p (t) . \label{eq:qUniform}
\end{equation}
Thus, the flow can be calculated from the instantaneous volume and conductance (the coefficient in front of $\Delta p$). In principle, both $r_s$ and $l_s$ can be imposed, but in an elastic material, these two functions are coupled. What remains is to consider physically reasonable choices of $r_s$ and $l_s$. We will consider two simple limits that demonstrate how mechanics couples $r_s$ and $l_s$.

\subsection{Uniform radius-imposed contractions}

\begin{figure}
    \centering
    \includegraphics[width=.8\linewidth]{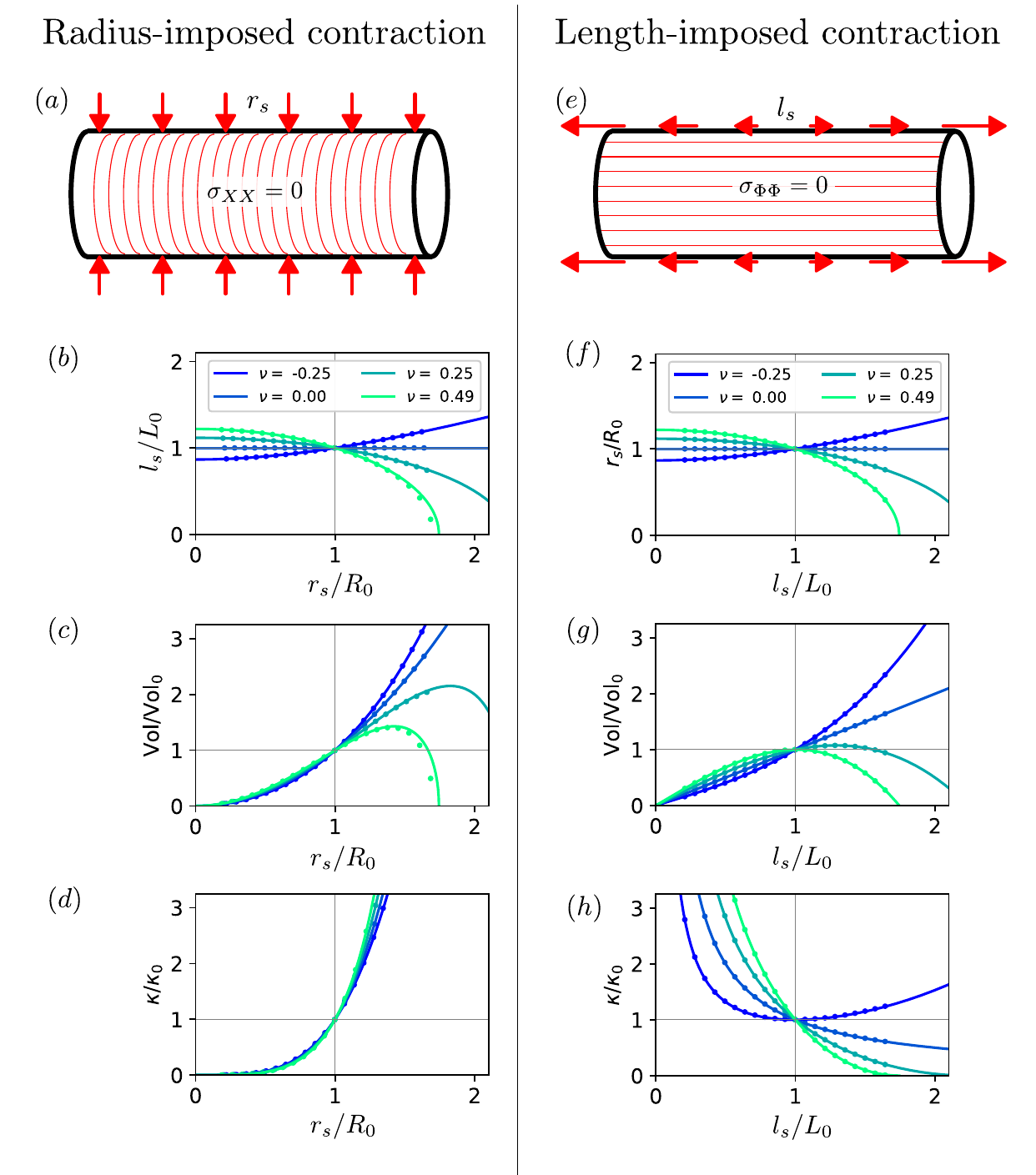}
    \caption{Uniform contractions in a finite tube of rest radius $R_0$ and rest length $L_0$. Two types of boundary conditions are considered: $(a)-(d)$ radius-imposed contraction and $(e)-(h)$ length-imposed contraction. $(a)$ During radius-imposed contraction, circumferential stresses (represented by red lines around the circumference) impose a radius of $r_s = R_0 + u_r$ on the tube, while the condition that $\sigma_{XX}=0$ determines the length of the tube according to \eqref{eq:dudx_radius_imposed}.     The results for the dimensionless $(b)$ length $l_s/L_0$, $(c)$ volume $\text{Vol}/\text{Vol}_0$, and $(d)$ conductance $\kappa/\kappa_0$ are plotted as a function of the applied radius. Different colors show different values of the Poisson ratio $\nu$. The solid curves show the analytical results \eqref{eq:ls_radial}, \eqref{eq:vol_radial}, and \eqref{eq:kappa_radial}, while points show results from COMSOL. $(e)$ During length-imposed peristalsis, longitudinal stresses (represented by red lines along the length) impose a length $l_s = L_0 + \frac{\partial u_x}{\partial X}L_0$, while the condition that $\sigma_{\Phi \Phi} = 0$ determines the radius fo the tube according to \eqref{eq:ur_length_imposed}. The results for the dimensionless $(f)$ radius $r_s/R_0$, $(g)$ volume $\text{Vol}/\text{Vol}_0$, and $(h)$ conductance $\kappa/\kappa_0$ are plotted as a function of the applied length.}
    \label{fig:UniformContraction}
\end{figure}

Suppose that the radius $r_s(t)$ is given at all times, but the length is determined by the condition that the stress $\sigma_{XX}$ vanishes. This could be the case in a scenario where  circumferential muscles contract a biological vessel, generating hoop stress while keeping the axial tension small. Then, equations \eqref{eq:epsilon_PHIPHI}, \eqref{eq:epsilon_XX}, and \eqref{eq:sigma_XX} give
\begin{equation}
    \sigma_{XX} = 0 \implies \epsilon_{XX} = - \nu \epsilon_{\Phi \Phi} \implies \frac{\partial u_x}{\partial X} = -1 + \sqrt{1-2\nu \left[\left( \frac{u_r}{R_0}\right)+\frac{1}{2}\left( \frac{u_r}{R_0}\right)^2 \right]}. \label{eq:dudx_radius_imposed}
\end{equation}
We will refer to this particular set of boundary conditions as ``radius-imposed contractions". An illustration is shown in figure \ref{fig:UniformContraction}$(a)$.

There is a maximum displacement that ensures that the tube length remains positive:
\begin{equation}
\frac{u_r}{R_0} \in 
    \begin{cases}
         \big[-1, -1+\sqrt{1+\nu^{-1}}\big], & \nu > 0\\ \big[-1, \infty \big), & \nu \leq 0
    \end{cases} .\label{eq:max_amplitude}
\end{equation}
For $\nu=0.5$, the max amplitude is 0.732. There is no positive real solution for the tube length corresponding to a tube which has expanded radially beyond that size. In these very large strain regimes, Hooke's law may no longer be valid, and a more appropriate nonlinear elastic theory should be used, but for the sake of this work that is focusing on geometry, this is a valid limit to take. 

For a tube of rest length $L_0$ and rest radius $R_0$ undergoing uniform deformation, if the radius $r_s$ is imposed, then the length $l_s$, volume $\text{Vol}$, and conductance $\kappa$ are 
\begin{equation}
    \frac{l_s}{L_0} = \sqrt{1-2\nu \left[\left( \frac{r_s-R_0}{R_0}\right)+\frac{1}{2}\left( \frac{r_s-R_0}{R_0}\right)^2 \right]} \label{eq:ls_radial},
\end{equation}
\begin{equation}
    \frac{\text{Vol}}{\text{Vol}_0} \equiv \frac{\pi r_s^2 l_s}{\pi R_0^2 L_0} = \frac{ r_s^2}{R_0^2} \sqrt{1-2\nu \left[\left( \frac{r_s-R_0}{R_0}\right)+\frac{1}{2}\left( \frac{r_s-R_0}{R_0}\right)^2 \right]} \label{eq:vol_radial},
\end{equation}
\begin{equation}
    \frac{\kappa}{\kappa_0} \equiv \frac{\pi r_s^4/8\mu l_s}{\pi R_0^4/8\mu L_0} = \frac{r_s^4}{R_0^4} \left[\sqrt{1-2\nu \left[\left( \frac{r_s-R_0}{R_0}\right)+\frac{1}{2}\left( \frac{r_s-R_0}{R_0}\right)^2 \right]}\right]^{-1} \label{eq:kappa_radial}.
\end{equation}
The results are plotted with solid lines in figure \ref{fig:UniformContraction}$(b)-(d)$. The points are obtained using COMSOL. Details of the COMSOL simulations are given in the appendix. The case $\nu=0.49$ corresponds to an almost incompressible solid while $\nu=0$ corresponds to the case where the length does not change during contraction (the typical case considered for contraction-induced flows \cite{Shapiro_Jaffrin_Weinberg_1969, Li1993Non-SteadyTubes}). No finite change in radius will cause the length to diverge to infinity, but a finite change in radius may cause the length to approach zero only if $\nu>0$ or $\nu = -1$. If $\nu > 0$, then the tube approaches zero length when $r_s$ approaches the maximum value allowed by \eqref{eq:max_amplitude}. As $\nu \rightarrow -1$, the length approaches zero as the radius approaches zero. For all other values of $\nu$, the length can never reach zero. In all cases, the volume reaches zero for $r_s=0$. For $r_s$ close to 1, an incompressible solid produces smaller volume changes than a tube of constant length when undergoing radial contractions. This suggests that the instantaneous flow induced by contractions in an incompressible solid free to stretch longitudinally may be overestimated if one neglects length changes as is often done. For very large increases in $r_s$ and $\nu>0$, a large increase in radius can actually decrease the volume since the length becomes so small (see the light green curve in panel $(c)$). The Poisson ratio has only a small effect on the conductance, as seen in panel $(d)$, since the factor of $r_s^4$ is the dominant term and is the same in all cases. 

\subsection{Uniform length-imposed contractions}

Now, suppose that the length $l_s(t)$ is given at all times, but the radius is determined by the condition that the stress $\sigma_{\Phi \Phi}$ vanishes. This could be the case in a scenario where longitudinal muscles contract a biological vessel, generating axial tension while keeping the hoop stress small. Then, \eqref{eq:sigma_PHIPHI} gives
\begin{equation}
    \sigma_{\Phi \Phi} = 0 \implies \epsilon_{\Phi \Phi} = - \nu \epsilon_{XX} \implies \frac{ u_r}{R_0} = -1 + \sqrt{1-2\nu \left[\left( \frac{\partial u_x}{\partial X}\right)+\frac{1}{2}\left( \frac{\partial u_x}{\partial X}\right)^2 \right]}. \label{eq:ur_length_imposed}
\end{equation}
We will refer to this particular set of boundary conditions as ``length-imposed contractions". An illustration is shown in figure \ref{fig:UniformContraction}$(e)$.

The results for length-imposed contractions are as follows:
\begin{equation}
    \frac{r_s}{R_0} = \sqrt{1-2\nu \left[\left( \frac{l_s-L_0}{L_0}\right)+\frac{1}{2}\left( \frac{l_s-L_0}{L_0}\right)^2 \right]} \label{eq:rs_longitudinal},
\end{equation}
\begin{equation}
    \frac{\text{Vol}}{\text{Vol}_0} = \frac{\pi r_s^2 l_s}{\pi R_0^2 L_0} = \frac{ l_s}{L_0} \left[1-2\nu \left[\left( \frac{l_s-L_0}{L_0}\right)+\frac{1}{2}\left( \frac{l_s-L_0}{L_0}\right)^2 \right]\right] \label{eq:vol_longitudinal},
\end{equation}
\begin{equation}
    \frac{\kappa}{\kappa_0} = \frac{\pi r_s^4/8\mu l_s}{\pi R_0^4/8\mu L_0} = \frac{L_0}{l_s} \left[1-2\nu \left[\left( \frac{l_s-L_0}{L_0}\right)+\frac{1}{2}\left( \frac{l_s-L_0}{L_0}\right)^2 \right]\right]^{2} \label{eq:kappa_longitudinal}.
\end{equation}
The results are plotted in figure \ref{fig:UniformContraction}$(f)-(h)$. The functional form of $r_s(l_s)$ is the same as $l_s(r_s)$ in the previous section, though the consequences of this relationship are different. One striking feature of length-imposed contractions is that the volume has a maximum at $l_s=L_0$ for $\nu = 1/2$ (see the light green curve in panel $(g)$), so a small-amplitude length change will lead to very little instantaneous flow in an incompressible vessel. Unsurprisingly, the factor of $r_s^4$ has a large effect on the conductance, so the conductance is sensitive to the Poisson ratio, as seen in panel $(h)$. At $\nu=-1/4$, small changes in the length from the rest length give only quadratic corrections to the conductance. For $\nu>0$, the conductance falls to zero at the maximum amplitude as the radius of the vessel constricts, but for $\nu < 0$, the conductance never falls to zero unless $\nu=-1$ and $l_s = 0$. Comparing figure \ref{fig:UniformContraction}$(d)$ and $(h)$, the conductance for length-imposed contractions looks completely different from that of radius-imposed contractions. We will see the full consequences of this form of the conductance when considering peristaltic waves.

\section{Peristaltic waves in an elastic tube }

\subsection{General properties of peristaltic waves}
Assume the tube boundary deforms as a wave-train with wavelength $\lambda$, period $T$, and speed $c = \lambda / T$ in a tube of infinite length. That is, 
\begin{align}
    \frac{\partial x_s(X,t)}{\partial X} &= \frac{\partial x_s(X-ct)}{\partial X}\\
    r_s(X,t) &= r_s(X-ct).
\end{align}
The pressure drop per wavelength is a key parameter in characterizing peristaltic pumping, so we will define the dimensionless parameter 
\begin{equation}
    \Delta \bar{p}_{\lambda} = \frac{p(X+\lambda)-p(X)}{8c \mu \lambda / R_0^2}. \label{eq:Deltapbar_lambda}
\end{equation}
Note that a positive value of $\Delta \bar{p}_\lambda$ corresponds to an adverse pressure gradient. 

The $T$-averaged flow in the lab frame for any wave-like contraction can be obtained by simplifying equation \eqref{eq:q(X,t)_general}:
\begin{equation}
    \langle q \rangle = c\pi R_0^2 \left[\Big{\langle} \Big(\frac{r_s}{R_0}\Big)^2 \Big(\frac{\partial x_s}{\partial X}\Big) \Big{\rangle} -\frac{ \Delta \bar{p}_{\lambda}}{\langle (\frac{r_s}{R_0})^{-4} \big(\frac{\partial x_s}{\partial X} \big) \rangle }  -  \frac{\langle (\frac{r_s}{R_0})^{-2} \big(\frac{\partial x_s}{\partial X}\big)^2 \rangle }{\langle (\frac{r_s}{R_0})^{-4} \big(\frac{\partial x_s}{\partial X}\big) \rangle } \right]. \label{eq:general_mean_flow}
\end{equation}
When $\frac{\partial x_s}{\partial X} = 1$, we recover the familiar expression for a radius-imposed peristaltic wave in a tube of fixed length \cite{Shapiro_Jaffrin_Weinberg_1969}, but here, the factors of $\frac{\partial x_s}{\partial X}$ are new and account for the longitudinal response. For a purely radial wave with $\frac{\partial x_s}{\partial X} = 1$ and $\Delta \bar{p}_{\lambda} = 0$, the flow always travels in the direction of the peristaltic wave. However, for a purely longitudinal wave with $\frac{r_s}{R_0} = 1$ and $\Delta \bar{p}_\lambda = 0$, we have
\begin{equation}
    \langle q \rangle = - \frac{c\pi R_0^2}{\Big{\langle}  \frac{\partial x_s}{\partial X} \Big{\rangle}}  \left[ \Big{\langle} \left( \frac{\partial x_s}{\partial X} \right)^2 \Big{\rangle} - \Big{\langle}  \frac{\partial x_s}{\partial X} \Big{\rangle}^2 \right],
\end{equation}
which is always negative for $c>0$, so the flow always travels in the opposite direction of the peristaltic wave. 
Because a peristaltic pump typically operates against an adverse pressure gradient, one often speaks of the ``pumping range" \cite{Kalayeh_Xie_Brian_Fowlkes_Sack_Schultz_2023}, where both $\langle q \rangle$ and $\Delta \bar{p}_\lambda$ are non-negative, and thus peristalsis acts like a pump. Interestingly, for $\Delta \bar{p}_\lambda > 0$, a peristaltic pump with only longitudinal wall motion must propagate a wave down the pressure gradient in order to act like a functioning pump. 

\subsubsection{Flow is steady in the co-moving wave frame}

Consider the material coordinates in the co-moving (wave) frame $\tilde{\Omega}_0$ which moves at a speed $c$. Traditionally uppercase and lowercase letters are used to distinguish lab frame from co-moving frame, but since we are using this notation to distinguish material coordinates from Eulerian coordinates, we will instead use tildes to denote any quantities in the co-moving frame. The wave-frame material coordinates $\vec{\tilde{X}}$ are related to the lab-frame material coordinates $\vec{X}$ by a Galilean transformation:
\begin{equation}
    \tilde{R} = R, \hspace{.5cm} \tilde{\Phi} = \Phi, \hspace{.5cm}  \tilde{X} = X-ct, \label{eq:wave_X}
\end{equation}
\begin{equation}
    \tilde{v}^R = v^R, \hspace{.5cm }\tilde{v}^X = v^X-c.\label{eq:wave_v}
\end{equation}
The analysis for peristaltic waves is usually performed in the co-moving frame where the flow is steady and the particles follow the streamlines. This is still the case for our problem, though in our material coordinates, it is not the boundary motion which travels as a wave, it is the metric which travels as a wave: 
\begin{equation}
    \frac{1}{\sqrt{|g|}} \frac{\partial \sqrt{|g|}}{\partial t} = - \frac{c}{\sqrt{|g|}} \frac{\partial \sqrt{|g|}}{\partial X}. \label{eq:metric_wave}
\end{equation}
In $\tilde{\Omega}_0$, incompressible flow is recovered as a consequence of \eqref{eq:metric_wave} and \eqref{eq:continuity_material}:
\begin{equation}
    \frac{1}{\sqrt{|g|}}\frac{\partial}{\partial \tilde{R}} (\sqrt{|g|}\tilde{v}^R) + \frac{1}{\sqrt{|g|}}\frac{\partial}{\partial \tilde{X}} (\sqrt{|g|} \tilde{v}^X) = 0,
\end{equation}
which is just the $\vec{\tilde
X}$ divergence of $\vec{\tilde{v}}$. This allows us to utilize the wave frame stream function 
\begin{equation}
    \tilde{\psi}( \tilde{R},\tilde{X}) = \int_0^{2\pi} \int_0^{\tilde{R}} \tilde{v}^X \sqrt{|g|} d\tilde{R} d \tilde{\Phi},
\end{equation}
along which particles in the wave frame will travel. By convention, we set $\tilde{\psi}(0,\tilde{X})=0$. 
The wave-frame flow $\tilde{q}$ is defined as 
\begin{equation}
    \tilde{q} \equiv  \int_0^{2\pi} \int_0^{\tilde{R}} \tilde{v}^X \sqrt{|g|} d\tilde{R} d \tilde{\Phi} = \tilde{\psi}(R_0,X), \label{eq:qTilde_definition}
\end{equation}
is independent of $\tilde{X}$, and is given by
\begin{align}
    \tilde{q} =  c\pi R_0^2 \left[ -\frac{ \Delta \bar{p}_{\lambda}}{\langle (\frac{r_s}{R_0})^{-4} \big(\frac{\partial x_s}{\partial \tilde{X}} \big) \rangle }  -  \frac{\langle (\frac{r_s}{R_0})^{-2} \big(\frac{\partial x_s}{\partial \tilde{X}}\big)^2 \rangle }{\langle (\frac{r_s}{R_0})^{-4} \big(\frac{\partial x_s}{\partial \tilde{X}}\big) \rangle } \right].
\end{align}
The first term can be thought of as pressure-driven flow (with a conductance that depends on the geometry), and the second-term is a nonlinear coupling between conductance and volume change. Throughout our analysis, we will assume that $\tilde{q}<0$ which will be the case unless a sufficiently large favorable pressure is applied. It immediately follows from the definitions of $\tilde{q}$, $q(X,t)$, and $\tilde{v}^X$ that 
\begin{equation}
    q(X,t) = \tilde{q} + c\pi r_s^2 \frac{\partial x_s}{\partial X}. \label{eq:q(X,t)_wave}
\end{equation}
The lab-frame flow $q(X,t)$ depends on $X$ and $t$ and may be positive or negative. 

The wave-frame stream function and velocity fields can easily be expressed in terms of $\tilde{q}$:
\begin{align}
    \tilde{\psi}(\tilde{R}, \tilde{X}) = \tilde{q} \frac{\tilde{R}^2}{R_0^2} \left( 2 - \frac{\tilde{R}^2}{R_0^2}\right) + c \pi r_s^2 \frac{\partial x_s}{\partial \tilde{X}} \frac{\tilde{R}^2}{R_0^2} \left( 1 - \frac{\tilde{R}^2}{R_0^2} \right),  \label{eq:psiTilde}
\end{align}
\begin{align}
    \tilde{v}^X(\tilde{R},\tilde{X}) &= 2c \left( 1 + \frac{\tilde{q}}{c\pi r_s^2 \frac{\partial x_s}{\partial \tilde{X}}}\right) \left( 1 - \frac{\tilde{R}^2}{R_0^2}\right) -c,\\
    \tilde{v}^R(\tilde{R},\tilde{X}) &= -\frac{c}{\pi r_s^2 \frac{\partial x_s}{\partial \tilde{X}}} \frac{\partial}{\partial \tilde{X}} \left(\pi r_s^2 \frac{\partial x_s}{\partial \tilde{X}} \right) \frac{\tilde{R}}{2}\left(1 - \frac{\tilde{R}^2}{R_0^2} \right).
\end{align}
Because these functions are time-independent, it is easy to calculate the wave-frame particle trajectories $\tilde{X}_p(t)$ and $\tilde{R}_p(t)$ by simultaneously integrating the equations:
\begin{equation}
    \frac{d\tilde{X}_p}{dt} = \tilde{v}^X(\tilde{R}_p,\tilde{X}_p), \hspace{1cm} \frac{d\tilde{R}_p}{dt} = \tilde{v}^R(\tilde{R}_p,\tilde{X}_p). \label{eq:dXTildepdt}
\end{equation}
Certain properties of peristaltic pumping, such as reflux, are only revealed by studying particle trajectories. An example velocity field in $\tilde{\Omega}_0$ is plotted in the right column of figure \ref{fig:threeCoordinateSystems}. Notice the velocity field is time-independent. In practice, all quantities are first calculated in $\tilde{\Omega}_0$ where flow is steady, then calculated in $\Omega_0$ using the Galilean transformation in equations \eqref{eq:wave_X} and \eqref{eq:wave_v}, and finally calculated in $\Omega(t)$ using equations \eqref{eq:xSigma} and \eqref{eq:Euler_V}. Note that, unlike the uniform contraction case where a simple coordinate inversion map exists, there is no analytic form for the velocity field, flow, or stream function as a function of the Eulerian coordinates in $\Omega(t)$. So, it is challenging to plot $\vec{V}(\vec{x},t)$ at a particular value of $x$, but one can instead plot $\vec{V}(\vec{x}_\Sigma(\vec{X},t),t)$ at a particular $\vec{x}_\Sigma(\vec{X},t)$, tracking points on the surfaces $\Sigma(X,t)$. Particle trajectories are found in a similar way. First, particle trajectories in $\tilde{\Omega}_0$ are obtained by integrating equation \eqref{eq:dXTildepdt}, then the coordinate transformations \eqref{eq:wave_X} and \eqref{eq:xSigma} are used to obtain particle trajectories in $\Omega_0$ and $\Omega(t)$. The velocity field, particle trajectory, and stream function plots (figures \ref{fig:threeCoordinateSystems}, \ref{fig:radialVelocityField}, \ref{fig:radialStreamlines}, \ref{fig:longitudinalVelocityField}, and \ref{fig:longitudinalStreamlines}) were all created in Mathematica using this approach.

\subsubsection{Trapping}
Individual particles typically travel much slower than the peristaltic wave speed, but at large amplitude, particles are trapped in a bolus and transported at average speed $c$ in the lab frame \cite{Shapiro_Jaffrin_Weinberg_1969}. During trapping, particles traverse closed streamlines encircling a fixed point in $\vec{\tilde{v}}$ in a region of $\tilde{\Omega}_0$ bounded by two curves $\tilde{R}(\tilde{X})$ solving $\tilde{\psi}(\tilde{R}, \tilde{X}) = 0$. For a visualization, see the dark blue regions in panels $(c)-(e)$ of figure \ref{fig:radialStreamlines}. Outside the trapping region, $\tilde{\psi}<0$, and inside the trapping region, $\tilde{\psi}>0$. Inverting equation \eqref{eq:psiTilde} gives
\begin{equation}
    \tilde{R}(\tilde{\psi},\tilde{X})^2 = \frac{(2\tilde{q}+c\pi r_s^2 \frac{\partial x_s}{\partial \tilde{X}}) \pm \sqrt{(2\tilde{q}+c\pi r_s^2 \frac{\partial x_s}{\partial \tilde{X}})^2 - 4 \tilde{\psi}(\tilde{q}+c\pi r_s^2 \frac{\partial x_s}{\partial \tilde{X}})}}{2(\tilde{q}+c\pi r_s^2 \frac{\partial x_s}{\partial \tilde{X}})} R_0^2.
\end{equation}
Outside the trapping region, only the $+$ solution corresponds to a real value of $\tilde{R}$, but inside the trapping region, both solutions are real and form closed loops.

The easiest way to test for trapping is to search for fixed points in the wave frame velocity field $\vec{\tilde{v}}(\vec{\tilde{X}})$. Along the centerline $\tilde{R}=0$, fixed points appear where $\tilde{q}=-\frac{1}{2}c\pi r_s^2 \frac{\partial x_s}{\partial \tilde{X}}$. Away from the centerline, fixed points appear at a relative extrema of $\pi r_s^2 \frac{\partial x_s}{\partial \tilde{X}}$. At least one fixed point exists whenever 
\begin{equation}
    \tilde{q} \geq \min_{\tilde{X}} \left(-\frac{1}{2}c\pi r_s^2 \frac{\partial x_s}{\partial \tilde{X}}\right). \label{eq:trapping_criteria}
\end{equation}
The typical picture is that at small amplitude, there are no fixed points in the velocity field, and above a critical amplitude, there are two saddles on the $\tilde{R}=0$ line and one center at $\tilde{R} >0$ around which solutions orbit. See figure \ref{fig:radialStreamlines}$(c)$, for example. But, we also observe cases with two fixed points, like in figure \ref{fig:radialStreamlines}$(d)$.

\subsubsection{Reflux}
Reflux is the phenomenon where particles close to the tube wall are transported against the flow direction. As clarified in \cite{1971RefluxVelocity}, the condition for reflux is a purely Lagrangian notion that requires finding the mean particle speed, as opposed to simply looking at the fluid velocity. The mean particle speed can be calculated from the particle trajectories and is uniquely determined by the wave-frame stream function $\tilde{\psi}$. Denoting the time it takes for a particle to travel one wavelength along a streamline in the wave frame as $T_p$, then the mean particle speed in the wave frame is $\tilde{s}_p \equiv -\lambda/T_p$. If the particle is in the trapping regime, it will never traverse a wavelength in the wave frame and has zero average velocity: $\tilde{s}_p=0$, as it travels in a bolus moving with the wave speed. Outside the trapping regime, 
\begin{equation}
    T_p(\tilde{\psi}) = \int_\lambda^0 \frac{d\tilde{X}}{\tilde{v}^X(\tilde{R}(\tilde{\psi},\tilde{X}),\tilde{X})}\implies \tilde{s}_p(\tilde{\psi}) \equiv -\frac{\lambda}{T_p(\tilde{\psi})} = \left[ \frac{1}{\lambda} \int_0^\lambda \frac{d\tilde{X}}{\tilde{v}^X(\tilde{R}(\tilde{\psi},\tilde{X}),\tilde{X})}\right]^{-1}.
\end{equation}
Thus, the mean particle speed in the wave frame is the harmonic mean of $\tilde{v}^X$. The mean particle speed in the lab frame is $s_p(\tilde{\psi}) = \tilde{s}_p(\tilde{\psi}) + c$. 

Reflux occurs when some particles move against the wave direction ($s_p < 0$) even if there is net flow in the wave direction ($\langle q \rangle > 0$). For example, the particles close to the wall in panels $(e)$ and $(f)$ of figure \ref{fig:radialVelocityField} are traveling to the left despite the rightward flow, so we say the particles are undergoing reflux. Following the calculation in Shapiro, et al. \cite{Shapiro_Jaffrin_Weinberg_1969}, because reflux is a phenomenon that occurs close to the wall boundary, we can characterize the condition for trapping by performing a perturbative expansion in $\tilde{\psi}-\tilde{q} > 0 $. Generalizing the result of Shapiro, et al. to the case with longitudinal displacements, reflux occurs whenever
\begin{equation}
    \tilde{q} < - c\pi \frac{\langle r_s^{-2} \big(\frac{\partial x_s}{\partial \tilde{X}}\big)^{-1} \rangle}{\langle r_s^{-4} \big(\frac{\partial x_s}{\partial \tilde{X}}\big)^{-2}\rangle}. \label{eq:reflux_criteria}
\end{equation}
When $\frac{\partial x_s}{\partial \tilde{X}}=1$, the right-hand side is precisely $\tilde{q}(\Delta p = 0)$. It is well known that for the case with zero longitudinal displacement, in an axisymmetric geometry, trapping is generic: An arbitrarily small adverse pressure gives rise to reflux \cite{Shapiro_Jaffrin_Weinberg_1969}. This is no longer the case when longitudinal effects are taken into account, so it will be necessary to study the effect of longitudinal wall motion on reflux. It was concluded by Kalayeh, et al. that longitudinal wall motion can suppress reflux in a 2D geometry \cite{Kalayeh_Xie_Brian_Fowlkes_Sack_Schultz_2023} . It remains to be seen what the role of reflux is in a 3D axisymmetric geometry which, even neglecting longitudinal wall motion, has very different refluxing behavior than in 2D.

Once the particle speed is known for each $\tilde{\psi}$, one can partition the streamlines into those which have $s_p >0$ and $s_p<0$ and define the following quantities:
\begin{equation}
    \langle q^+ \rangle = \int_{s_p>0} v^X \sqrt{g} dR d\Phi, \label{eq:qPlus}
\end{equation}
\begin{equation}
    \langle q^- \rangle = \int_{s_p<0} v^X \sqrt{g} dR d\Phi. \label{eq:qMinus}
\end{equation}
The integrals are taken over all coordinates such that $s_p$ is positive or negative, respectively. Since $\langle q \rangle = \langle q^+ \rangle + \langle q^- \rangle$, this decomposes the flow into forward flow and backward flow. Similar definitions were used in \cite{Kalayeh_Xie_Brian_Fowlkes_Sack_Schultz_2023}.

\subsection{Radius-imposed peristaltic waves}

Consider waves of the form:
\begin{align}
    \frac{r_s}{R_0} &= 1 + \epsilon \cos (2\pi \tilde{X}) \label{eq:rs_RadialWave} \\
    \frac{\partial x_s}{\partial \tilde{X}} &= \sqrt{1-2\nu \left[\epsilon \cos (2\pi \tilde{X}) + \frac{1}{2} \epsilon^2 \cos^2 (2\pi \tilde{X})  \right]} \label{eq:dxsdX_RadialWave}
\end{align}
where the latter is chosen to enforce $\sigma_{XX}=0$ for a given $\nu$ according to equation \eqref{eq:dudx_radius_imposed}. 

One interesting property of these boundary conditions is that the Eulerian distance between two identical points on the wave: $\int_0^\lambda \frac{\partial x_s}{\partial \tilde{X}} d\tilde{X}$ is not necessarily equal to $\lambda$. For materials with $\nu>0$, wave-like contractions induce a global shortening. For example, a material with $\nu=0.5$ undergoing radial contractions at the maximum amplitude allowed by \eqref{eq:max_amplitude}, has its length reduced to 85.6 percent its original size, despite undergoing radial expansion and radial contraction in equal amounts.

\begin{figure}
    \centering
    \includegraphics[width=\linewidth]{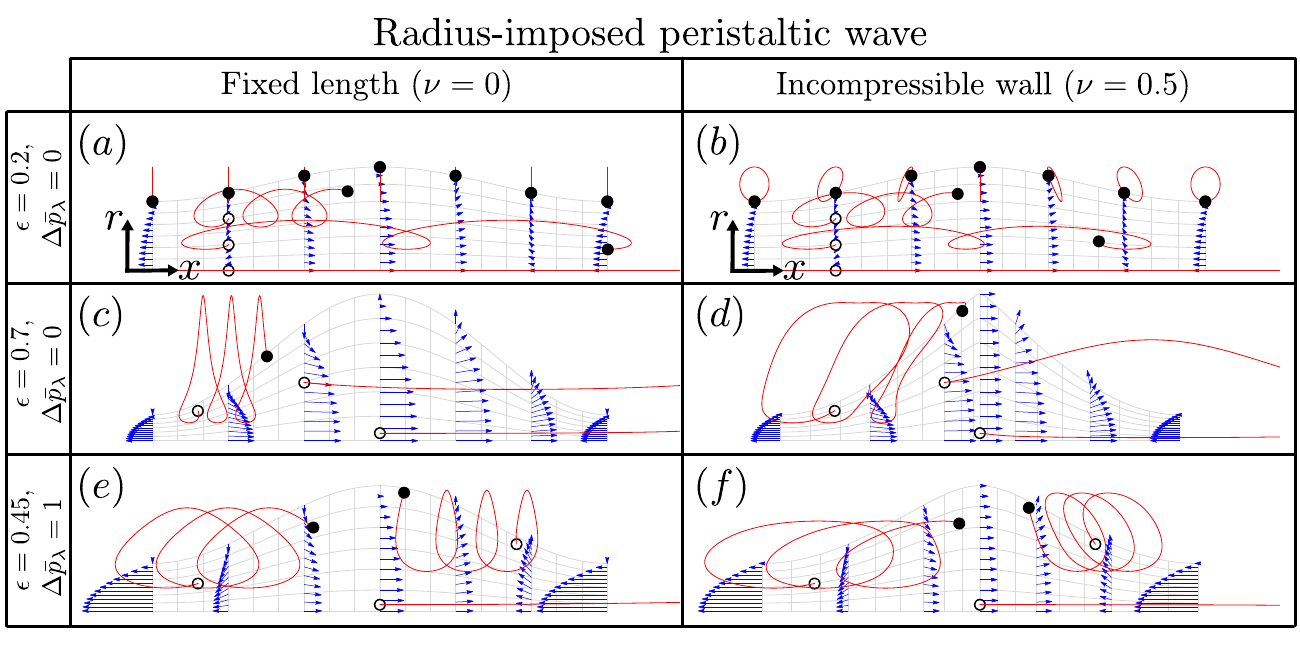}
        \caption{Instantaneous Eulerian velocity field $\vec{V}(\vec{x},0)$ in the lab frame $\Omega(t=0)$ and example particle trajectories $\vec{x}_p(t)$ for a tube driven by a radius-imposed peristaltic wave traveling to the right. Solutions for a purely radial wall motion $\nu=0$ and an incompressible tube with $\sigma_{XX}=0$ are compared. Each row corresponds to different choices of the characteristic strain $\epsilon$ and adverse pressure $\Delta \bar{p}_\lambda$. Gridlines which are equally spaced for a tube at rest help to visualize longitudinal displacements. The length of the arrows indicates the magnitude and direction of $\vec{V}$. Particles begin their trajectories at $t=0$ at the locations marked by hollow circles and end their trajectories at $t=3T$ at the locations marked by filled-in circles. }
    \label{fig:radialVelocityField}
\end{figure}

\begin{figure}
    \centering
    \includegraphics[width=\linewidth]{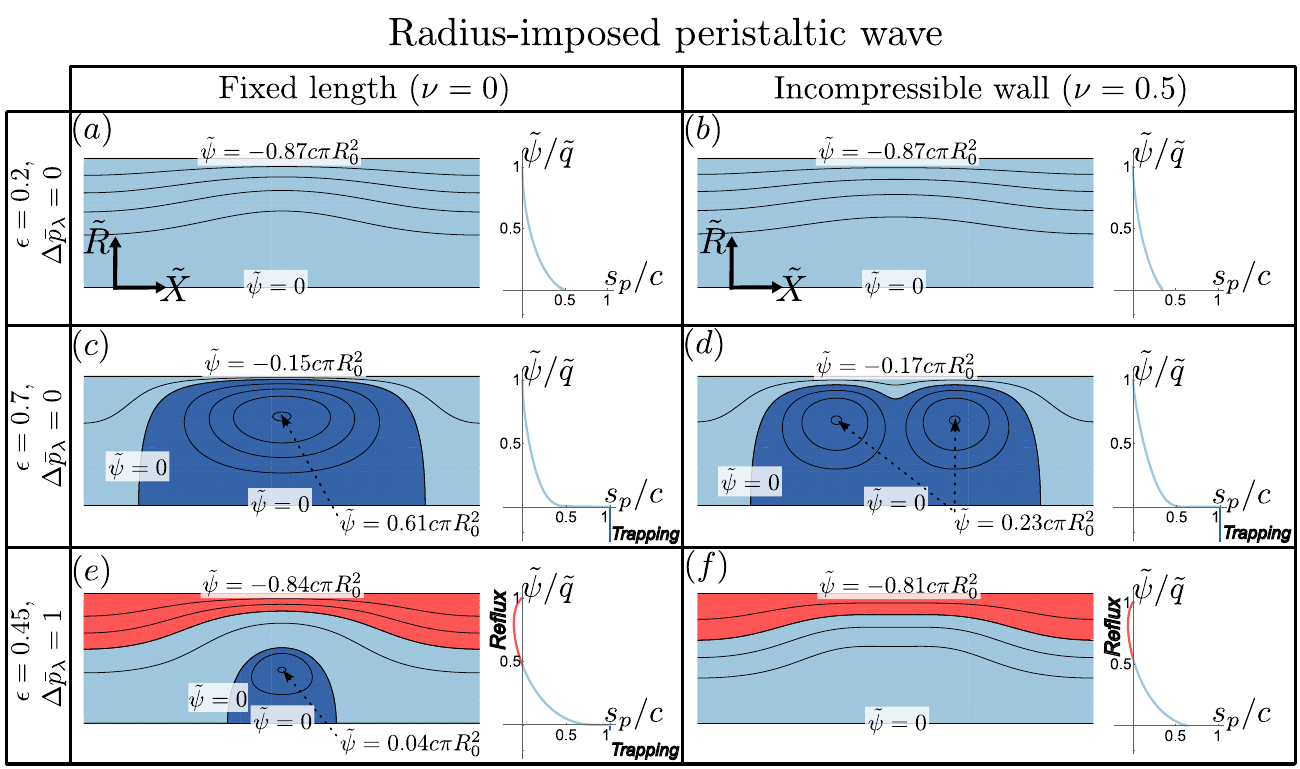}
    \caption{Streamlines in the co-moving wave-frame $\tilde{\Omega}_0$ and the corresponding particle wave speed distribution are plotted for a tube driven by a radius-imposed peristaltic wave using the same parameters as in figure \ref{fig:radialVelocityField}. At $\tilde{R}=0$, $\tilde{\psi}=0$ by convention. At $\tilde{R}=R_0$, $\tilde{\psi} = \tilde{q}$; its value is noted. Positive values of $\tilde{\psi}$ (negative values of $\tilde{\psi}/\tilde{q}$) indicate trapping while negative values of $s_p$ indicate reflux. For visualization, regions where particles undergo trapping are colored dark blue, regions where particles undergo reflux are colored red, and regions where particles travel forward at a speed less than $c$ are colored light blue.}
    \label{fig:radialStreamlines}
\end{figure}

Some example velocity fields and particle trajectories are shown in figure \ref{fig:radialVelocityField}. A comparison is made between the standard peristalsis calculation $\nu = 0$ and a more realistic incompressible material $\nu = 0.5$. The lab-frame velocity field $\vec{V}(\vec{x})$ in $\Omega(t=0)$ is shown, which is related to $\vec{v}(\vec{X})$ by equation \eqref{eq:Euler_V}. When $\nu = 0$, particles on the wall are displaced in the radial direction only, as seen in panel $(a)$, but when $\nu = 0.5$, particles on the wall sweep out loops, as seen in panel $(b)$. The velocity profile $V^x(r)$ is in general a sum of a parabolic term coming from $v^X(R)$ and a constant in $R$ term $\frac{\partial x_s}{\partial t}$, though the latter vanishes for $\nu = 0$. At small amplitude (panels $(a)$ and $(b)$), the velocity field and particle trajectories look similar for the $\nu = 0$ and $\nu = 0.5$ cases. At larger amplitude, the flat profile is visible in the incompressible tube (panel $(d)$), especially where the radius is large, forcing the local length to decrease in order to conserve tube volume. In all cases, a positive value of $V^x$ is observed in the expanded regions where the factor of $c\pi r_s^2 \frac{\partial x_s}{\partial X}$ in \eqref{eq:q(X,t)_wave} is large, and a negative value of $V^x$ is observed in the contracted regions where $c\pi r_s^2 \frac{\partial x_s}{\partial X}$ is small.

Particle trajectories in figure \ref{fig:radialVelocityField} are shown over a time of three wave periods. When an adverse pressure is applied (panels $(e)$ and $(f)$), refluxing particles can be observed close to the tube wall for both the fixed-length tube and the incompressible tube. For all parameters, the particles trace out seemingly complex trajectories in the lab frame. 

The wave-frame streamlines are plotted in figure \ref{fig:radialStreamlines}. The value of $\tilde{\psi}$ at $\tilde{R} = R_0$ is exactly the wave-frame flow $\tilde{q}$, and its value is noted at the top of each panel. Particles in $\tilde{\Omega}_0$ follow the streamlines, and the complex paths seen in $\Omega(t)$ can be understood as a coordinate transformation of these much simpler paths in $\tilde{\Omega}_0$. In addition to the streamlines, each panel shows the mean particle speed $s_p/c$ as a function of $\tilde{\psi}/\tilde{q}$, though the axes are inverted to suggest that a change in $\tilde{\psi}/\tilde{q}$ is typically a change in the vertical coordinate $\tilde{R}$; this convention has been used for other studies of peristalsis \cite{Shapiro_Jaffrin_Weinberg_1969, Takabatake_Ayukawa_Mori_1988}. These plots help to indicate the magnitude of particle trajectories along a particular streamline. At small amplitude, $\tilde{\psi}$ is always negative, and all particles travel rightward at a speed well below $c$. Light blue shading is used to indicate these non-trapped, non-refluxing particles. At large amplitude, trapping occurs, as indicated by positive values of $\tilde{\psi}$ or negative values of $\tilde{\psi}/\tilde{q}$ and visualized with dark blue shading. Interestingly, the incompressible material in panel $(d)$ has a pair of vortices. This behavior is only observed at very large amplitude, specifically in the regime where the local volume is maximized at multiple locations. Recall that this type of material has the peculiar property that the volume may decrease locally where the radius increases, see figure \ref{fig:UniformContraction}$(c)$. Typically, the bolus is centered at the maximum of $r_s$, but more generally, the bolus is centered at the maxima of $r_s^2 \frac{\partial x_s}{\partial \tilde{X}}$ which can occur at two locations for $\nu=0.5$ and $\epsilon$ sufficiently large. Note that the value of $\tilde{q}$ is similar to that of the material with fixed length, so its not clear that the double vortex enhances flow. When an adverse pressure gradient is introduced (panels $(e)$ and $(f)$), reflux can be observed as indicated by negative values of $s_p/c$ and colored in red. Interestingly, for this choice of parameters, trapping is observed for $\nu=0$ but not $\nu=0.5$. Panels $(c)-(f)$ demonstrate that changing the Poisson ratio can lead to topological changes in the flow field.

\begin{figure}
    \centering
    \includegraphics[width=\linewidth]{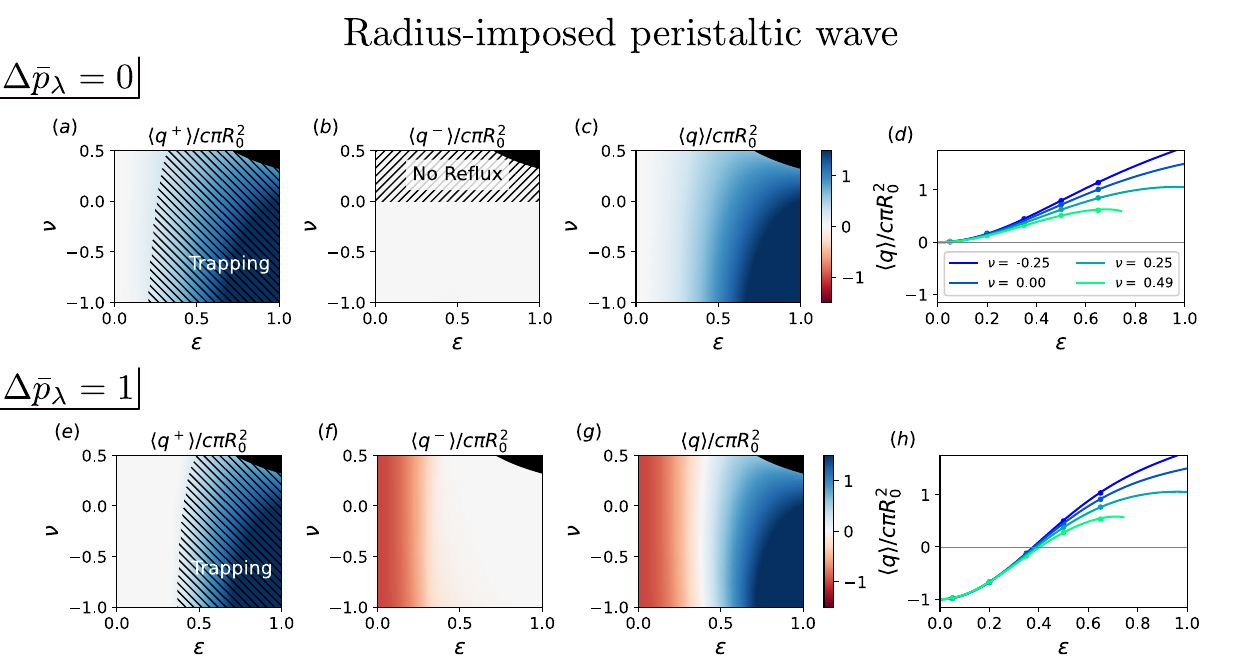}
    \caption{$(a)-(d)$ Flow for radius-imposed peristalsis with $\sigma_{XX}=0$ and zero pressure drop per wavelength for different choices of wave amplitude $\epsilon$ and Poisson ratio $\nu$. The colorbar is used for plots $(a)-(c)$ to indicate the strength of the flow with blue used for positive flow, red for negative flow, and white for zero flow. $(a)$ Mean flow over streamlines with positive particle speed, as given by equation \eqref{eq:qPlus}. The region where trapping occurs according to equation \eqref{eq:trapping_criteria} is hashed. $(b)$ Mean flow over streamlines with negative particle speed. Although all negative values of $\nu$ give nonzero backflow, the magnitude of the flow is negligible compared to that of the positive flow. $(c)$ Mean flow equal to the sum of the flows in $(a)$ and $(b)$. In all cases, the black regions in the top-right corner denote unphysical regions with negative tube length, according to equation \eqref{eq:max_amplitude}. $(d)$ The analytic mean flow at particular horizontal cuts of $(c)$ are shown (solid curves) and compared with COMSOL simulations (points). $(e)-(f)$ Repeated, but for adverse pressure $\Delta \bar{p}_{\lambda} = 1$. The pressure produces backflow which dominates the flow at small amplitude $\epsilon$, but once the amplitude is large enough for trapping to occur, the flow becomes positive again.}
    \label{fig:flow_radial_wave}
\end{figure}

The mean flow for arbitrary amplitudes can be obtained by substituting \eqref{eq:dudx_radius_imposed} into \eqref{eq:general_mean_flow}. At small amplitude, the solution is 
\begin{equation}
    \langle q \rangle = c\pi R_0^2 \left[-\left( 1 - (4+\nu) \Big\langle \frac{u_r}{R_0}\Big\rangle \right) \Delta \bar{p}_\lambda +  (8-2\nu - \nu^2)\left(\Big\langle \frac{u_r^2}{R_0^2}\Big\rangle  - \Big\langle \frac{u_r}{R_0}\Big\rangle^2 \right) \right].  \label{eq:q_radial_small}
\end{equation}
The contraction-induced flow is quadratic in the displacement amplitude and is largest for negative Poisson ratios. However, at small amplitude, the factors of $\nu$ provide only minor corrections to the case of zero longitudinal displacement $\nu=0$. 

At larger amplitudes, the solutions diverge and the difference is more pronounced. The mean flow for arbitrary amplitude when driven by a sine wave of amplitude $\epsilon$ is shown in figure \ref{fig:flow_radial_wave} for all values of $\nu$ and $\epsilon$ and two choices of $\Delta \bar{p}_\lambda$. The mean flow is further decomposed into positive and negative parts as given by equations \eqref{eq:qPlus} and \eqref{eq:qMinus}. When no adverse pressure exists in the tube (panels $(a)-(d)$), trapping occurs for $\epsilon$ approximately greater than 0.2, with a slightly larger regime of trapping for smaller values of $\nu$. Zero reflux is observed for $\nu>0$, but nonzero reflux occurs for $\nu < 0$. Typically reflux is thought to come about as a competition between peristalsis driving flow to the right and pressure driving flow to the left, but here there is only peristalsis, and the combination of radial and longitudinal effects leads to reflux. Though, for these negative Poisson ratios, $\langle q^-\rangle$ is still so much smaller in magnitude than $\langle q^+\rangle$ that it is unnoticeable in our colormap. When $r_s$ falls to zero, plug flow results such that $\tilde{q} = 0$, and the mean flow approaches the integrated volume divided by the period. For $\nu=0$, the integrated volume at $\epsilon = 1$ is $1.5$ times the tube's rest volume, and for negative values of $\nu$, it is even larger, but for $\nu=0.5$, it never even achieves plug flow since such large radius changes would be forbidden by \eqref{eq:max_amplitude}. At its maximum allowed amplitude, the incompressible tube generates a flow of only 0.576$c\pi R_0^2$, the smallest maximum flow of all Poisson ratios. Intuitively, the volume in an incompressible tube cannot get too large since an increase in radius will necessarily be accompanied by a decrease in length, while for an auxetic material $(\nu < 0)$, an increase in radius is accompanied by an increase in length, so the volume changes are enhanced.

When an adverse pressure $\Delta \bar{p}_{\lambda}=1$ exists in the tube (panels $(e)-(h)$), a large refluxing flow exists for all $\epsilon$ up until about the trapping transition where the flow is dominated by forward flow. 

In all cases, the analytic results agree well with the COMSOL simulations as shown with points in panels $(d)$ and $(h)$. The simulations were performed by imposing a radius on the exterior of a tube with small but finite thickness and $\sigma_{XR} = 0$ on the exterior of the tube. The Young's modulus was kept large enough that the fluid pressure exerts negligible force on the tube. These boundary conditions seem to reproduce the simple $\sigma_{XX} = 0$ condition used for analytic calculations. Thus, our lubrication approximation and finite strain calculations are valid for the parameters used in the simulation. See the appendix for more details.

Figure \ref{fig:flow_radial_wave} indicates that with or without adverse pressure, small changes in $\nu$ tend to lead to rather small changes in the flow at small to moderate amplitudes. The gradients in the colormaps in the $\epsilon$ direction of figure \ref{fig:flow_radial_wave} are much greater indicating that the amplitude has a greater effect than the Poisson ratio at setting the flow. Still, the Poisson ratio can lead to large differences in flow at large amplitude.

\begin{figure}
    \centering
    \includegraphics[width=0.72\linewidth]{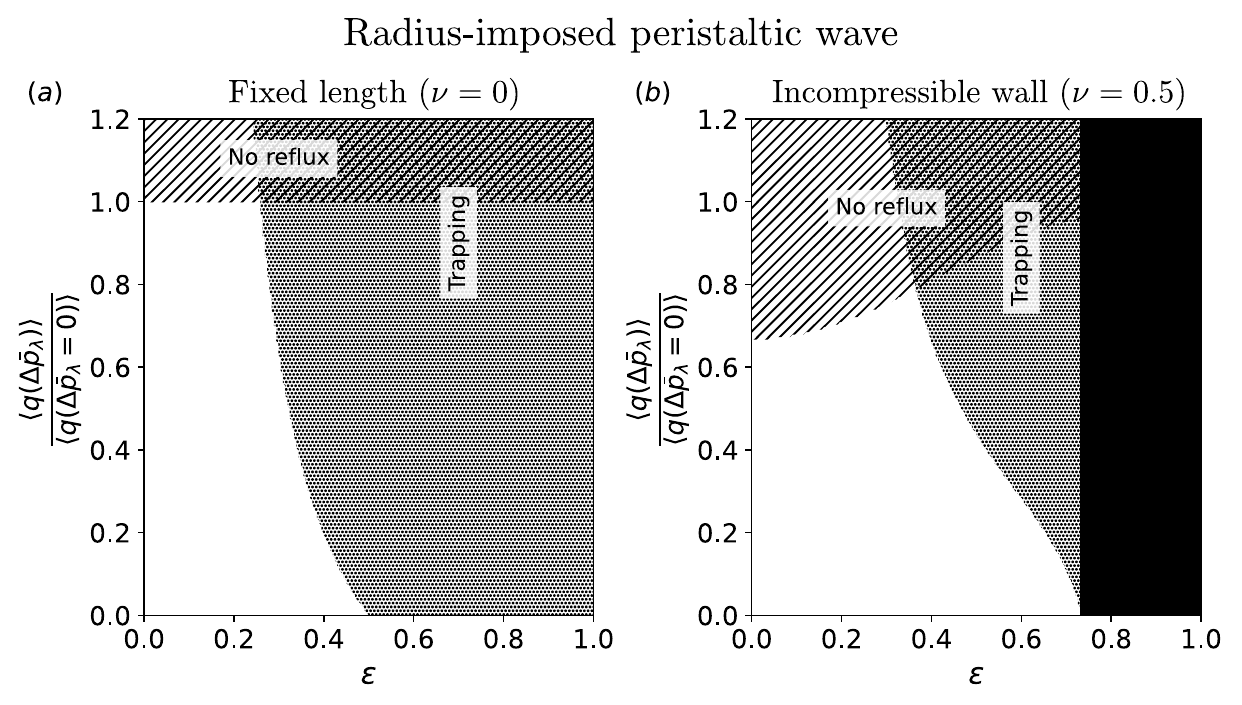}
    \caption{Phase diagram for radius-imposed peristalsis. The vertical axis shows the mean flow at a given pressure normalized by the mean flow at zero pressure. In $(a)$, the case with zero longitudinal displacement is shown, replicating the result in figure 13 of \cite{Shapiro_Jaffrin_Weinberg_1969}. In $(b)$, the case for an incompressible material is shown. By this measure, incompressibility can suppress reflux, but also can significantly suppress trapping. The black region on the right is the region forbidden by equation \eqref{eq:max_amplitude}. }
    \label{fig:phase_diagram}
\end{figure}

To further characterize reflux, it is convenient to consider the ratio $\langle q(\Delta \bar{p}_\lambda) \rangle / \langle q(\Delta \bar{p}_\lambda = 0) \rangle$ which takes values between 0 and 1 in the pumping regime (adverse pressure, positive flow). It was demonstrated in \cite{Shapiro_Jaffrin_Weinberg_1969} that for all values in the pumping regime, a radially contracted cylindrical tube with zero longitudinal displacement will always exhibit some reflux. This plot is re-created in figure \ref{fig:phase_diagram}$(a)$. However, for any other value of $\nu$, this is no longer the case. For $\nu=0.5$, there is a region within the pumping regime where reflux is completely suppressed, as shown in figure \ref{fig:phase_diagram}$(b)$. The no-reflux region is particularly large for small amplitudes. For the $\nu=0.5$ case, the trapping region is also smaller, particularly at low flow rates.

\subsection{Length-imposed peristaltic waves}
Now, consider waves of the form:
\begin{align}
    \frac{\partial x_s}{\partial \tilde{X}} &= 1 + \epsilon \cos (2\pi \tilde{X}) \label{eq:dxsdX_longitudinal} \\
     \frac{r_s}{R_0} &= \sqrt{1-2\nu \left[\epsilon \cos (2\pi \tilde{X}) + \frac{1}{2} \epsilon^2 \cos^2 (2\pi \tilde{X})  \right]} \label{eq:ur_longitudinal}
\end{align}
where the latter is chosen to satisfy the condition $\sigma_{\Phi \Phi}=0$, equation \eqref{eq:ur_length_imposed}.

The mean radial displacement is no longer zero. For materials with $\nu > 0$, longitudinal waves cause a global reduction in radius despite undergoing longitudinal stretching and squeezing in equal amounts.

\begin{figure}
    \centering
    \includegraphics[width=\linewidth,trim={0 3.28cm 0 0},clip]{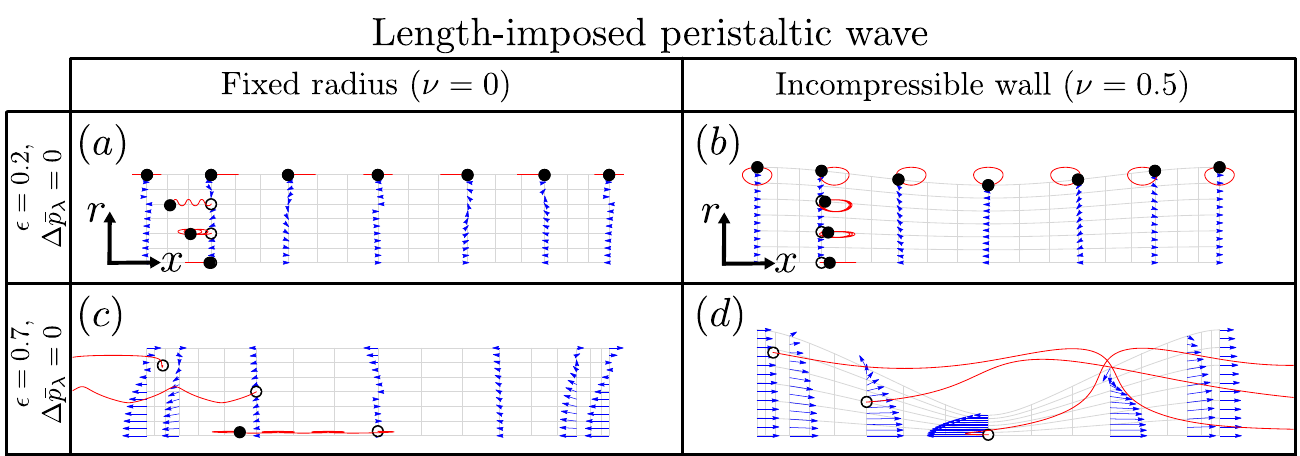}
        \caption{Instantaneous Eulerian velocity field $\vec{V}(\vec{x},0)$ in the lab frame $\Omega(t=0)$ and example particle trajectories $\vec{x}_p(t)$ for a tube driven by a length-imposed peristaltic wave traveling to the right. Solutions for a purely longitudinal wall motion $\nu=0$ and an incompressible tube with $\sigma_{\Phi \Phi}=0$ are compared. The two rows correspond to small and large amplitude, each at zero pressure. The length of the arrows is proportional to the magnitude of the velocity, with the same scaling factor used in figure \ref{fig:radialVelocityField} for comparison purposes. Particles begin their trajectories at $t=0$ at the locations marked by hollow circles and end their trajectories at $t=3T$ at the locations marked by filled-in circles. }
    \label{fig:longitudinalVelocityField}
\end{figure}

\begin{figure}
    \centering
    \includegraphics[width=\linewidth,trim={0 3.72cm 0 0},clip]
    {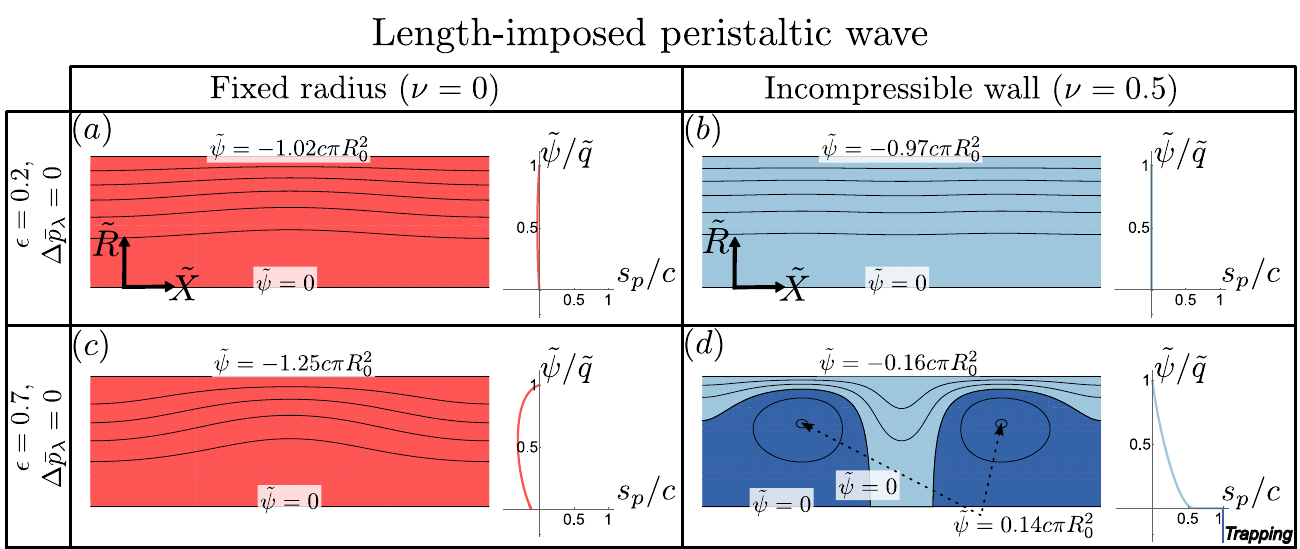}
    \caption{Streamlines in the co-moving wave-frame $\tilde{\Omega}_0$ plotted in material coordinates and the corresponding particle wave speed distribution are plotted using the same parameters as in figure \ref{fig:longitudinalVelocityField}. Regions where particles undergo trapping are colored dark blue, regions where particles undergo reflux are colored red, and regions where particles travel forward at a speed less than $c$ are colored light blue.}
    \label{fig:longitudinalStreamlines}
\end{figure}

Some sample velocity fields and particle trajectories are shown in figure \ref{fig:longitudinalVelocityField} at zero pressure, and the corresponding stream function and particle speeds are shown in figure \ref{fig:longitudinalStreamlines}. When $\nu = 0$, particles on the boundary are only displaced longitudinally, but when $\nu = 0.5$, particles sweep out closed loops. At small amplitude, both the $\nu = 0$ and $\nu=0.5$ materials lead to only tiny particle displacements after three wave periods. The $\nu=0.5$ material in particular leads to almost imperceptible particle displacement after three periods since $\langle q \rangle$ vanishes at order $\epsilon^2$, as calculated below. Recall that $V^x$ is the sum of a pressure gradient term which generates a parabolic profile $v^X(R)$ and a boundary term. Since the magnitude of $v^X(R)$ is always maximal at $R=0$, it can be that the profile $V^x(r)$ reverses direction: Close to the wall, the boundary term drives the velocity in one direction, but close to the centerline, the pressure gradient term drives the velocity in the opposite direction, as seen in both panels $(a)$ and $(c)$ of figure \ref{fig:longitudinalVelocityField}. In panel $(b)$, the boundary term dominates, so the profile is essentially flat. The scaling of arrow length with velocity magnitude is the same as that of figure \ref{fig:radialVelocityField}, so notice that the magnitude of the velocities in panels $(a)$ and $(b)$ in figure \ref{fig:longitudinalVelocityField} are much less than the equivalent plots in figure \ref{fig:radialVelocityField}. At larger amplitude, the effect of the Poisson ratio is more pronounced. For the $\nu=0$ material, all particles travel to the left, but for the $\nu=0.5$ material, two-vortex trapping occurs as seen in panel \ref{fig:longitudinalStreamlines}$(d)$, and all particles travel to the right. 

At small amplitude, the mean flow is
\begin{equation}
    \langle q \rangle = c\pi R_0^2 \left[-\left( 1 + (1+4\nu) \Big\langle \frac{\partial u_x}{\partial \tilde{X}}\Big\rangle \right) \Delta \bar{p}_\lambda +  (-1-2\nu + 8\nu^2)\left(\Big\langle \left(\frac{\partial u_x}{\partial \tilde{X}}\right)^2\Big\rangle  - \Big\langle \frac{\partial u_x}{\partial \tilde{X}}\Big\rangle^2 \right) \right].  \label{eq:q_longitudinal_small}
\end{equation}
The second term vanishes for $\nu = 1/2$ and $\nu=-1/4$. In the former case, the volume change vanishes to lowest order, and in the latter case, the conductance change vanishes so no pressure gradient is established when the tube is contracted, consistent with the observations for uniformly contracting tubes in figure \ref{fig:UniformContraction}. The fact that the contraction-induced flow vanishes to lowest order when $\nu=1/2$ suggests that small-amplitude length-imposed peristalsis is inefficient at pumping flow. A similar observation was made for a finite half-closed tube subject to an axial force \cite{Elbaz_Gat_2014}. For values of $\nu$ between $-1/4$ and $1/2$, the flow is negative. Only for $\nu<-1/4$ is the contraction-induced flow in the same direction as the traveling wave. Interestingly, the magnitude of the mean flow at small amplitude is maximized for a completely auxetic material ($\nu=-1$), as was the case for radius-imposed peristalsis. Only in the case $\nu = -1$ is the magnitude of the mean flow for a length-imposed peristaltic wave as big as the magnitude of the mean flow for a radius-imposed wave; in all other cases, the factor $(8-2\nu - \nu^2)$ appearing in equation \eqref{eq:q_radial_small} is larger in magnitude than the factor $(-1-2\nu + 8\nu^2)$ appearing in equation \eqref{eq:q_longitudinal_small}. It is worth stressing that the small-amplitude flow for either the radius-imposed or length-imposed peristaltic wave is order $\epsilon^2$, but the geometric prefactor leads to larger flow for radius-imposed peristalsis.

\begin{figure}
    \centering
   \includegraphics[width=\linewidth]{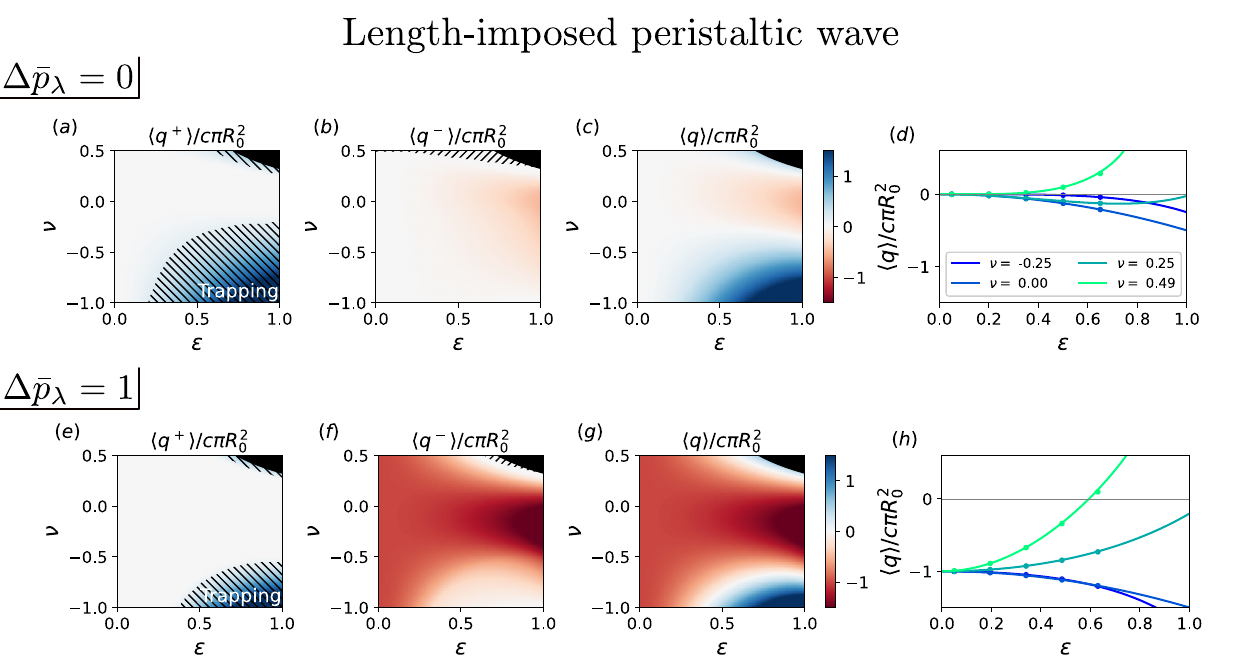}
    \caption{Flow for length-imposed peristalsis with $\sigma_{\Phi \Phi}=0$ and $(a)-(d)$ $\Delta \bar{p}_\lambda = 0$ or $(e)-(h)$ $\Delta \bar{p}_\lambda = 1$. Coloring conventions follow the ones of the plot for radial waves, figure \ref{fig:flow_radial_wave} (blue used for positive flow, red for negative flow, and white for zero flow). Mean flow over streamlines with $(a)$  positive particle speed, and $(b)$ negative particle speed. Panel $(c)$ is the net mean flow. $(d)$ The analytic mean flow at particular horizontal cuts of $(c)$ are shown (solid curves) and compared with COMSOL simulations (points). $(e)-(h)$ Repeated, but for adverse pressure $\Delta \bar{p}_{\lambda} = 1$. 
    Compared to figure \ref{fig:flow_radial_wave}, length-imposed peristalsis produces more negative flow and less positive flow. }
    \label{fig:flow_longitudinal_wave}
\end{figure}


At larger amplitude, an incompressible solid can generate positive flow when driven by longitudinal peristalsis. The mean flow results for all choices of $\epsilon$ and $\nu$ and two choices of $\Delta \bar{p}_{\lambda}$ are shown in figure \ref{fig:flow_longitudinal_wave}. The qualitative behavior for large-amplitude length-imposed peristalsis is very different from radius-imposed peristalsis. For radius-imposed peristalsis, at large amplitude, the conductance becomes very small at large values of $\epsilon$ where the radius approaches zero. For length-imposed peristalsis, conductance may or may not fall to zero at large amplitude. For example, when $\nu = 0.49$ and the length is stretched to its maximum value, the conductance reaches zero (see the green curve in figure \ref{fig:UniformContraction}$(h)$). However, when $\nu=-0.25$, the dimensionless conductance $\kappa/\kappa_0$ always remains greater than one (see the dark blue curve in figure \ref{fig:UniformContraction}$(h)$). When $\nu = -1$, the conductance vanishes when the length and radius simultaneously contract. For intermediate values of $\nu$ (not too close to $\nu = -1$ or $\nu = 0.5$), the radius never reaches zero with the sinusoidal driving considered here. Large positive flow and trapping are only observed for large-amplitude sinusoidal longitudinal waves with $\nu$ close to either $0.5$ or  $-1$, as shown in panels $(a)$ and $(c)$ of figure \ref{fig:flow_longitudinal_wave}. The disparity in the flow in these special regimes compared to intermediate values of $\nu$ becomes even more pronounced when an adverse pressure is applied as seen in panels $(e)-(h)$. 

\section{Networks of contracting vessels}
The analysis thus far is appropriate for studying tubular organs such as the esophagus, ureter, and embryonic heart, but contractions also partially or in some cases entirely drive flow through complex networks. The most famous example of this is the slime mold \textit{Physarum polycephalum} \cite{Alim2013RandomIndividual}. In general, the tools developed in this work could be relevant to all situations where a network of soft vessels is subject to large deformations, as for example is the coronary arteries and vein due to the contractility of the heart \cite{kurachi2000heart}. Here, we will briefly outline how to apply models of displacement-imposed peristalsis to networks and why the material coordinates used throughout the paper arise naturally in networks. 

Assume that each edge in the network undergoes a uniform contraction, and nodes are placed at the junctions of these edges. A node will be indexed with a Latin subscript $i,j,...$, and an edge will be indexed with a Greek subscript $\alpha, \beta, ...$, or a pair of Latin subscripts when it is necessary to specify the head and tail nodes that bound an edge. The oriented connectivity of the network is described by the incidence matrix whose elements $\Delta^T_{i\beta}$ are given by
\begin{equation}
    \Delta^T_{i\beta} = \begin{cases}
        +1, & \text{if edge $\beta$ enters node $i$}\\
        -1, & \text{if edge $\beta$ leaves node $i$}\\
        0, & \text{otherwise}\\
    \end{cases}. \label{eq:incidence}
\end{equation}

Revisiting equation \eqref{eq:qUniform}, the average of the flow across the tube length is directly related to the pressure drop across the end points of the tube at $X=-L_0/2$ and $X=+L_0/2$:
\begin{equation}
    Q(t) \equiv \int_{-L_0/2}^{L_0/2} q(X,t) \frac{dX}{L_0} = -\frac{\pi r_s(t)^4}{8\mu l_s(t)}  \left[p(L_0/2,t) - p(-L_0/2,t)\right] .\label{eq:qEdgeAvg}
\end{equation}
The quantity $Q(t)$ defines the edge flow, while $p(\pm L_0/2,t)$ describes the node pressures. Note that \eqref{eq:qUniform} implies that $Q(t) = q(0,t)$, so for a uniformly contracting tube, the edge flow is both the edge-averaged flow and the flow at the center of the tube. Applying equation \eqref{eq:qEdgeAvg} to an edge $\alpha$ in the network gives us an equation analogous to Ohm's law:
\begin{equation}
    Q_{\alpha}(t) = - \sum_j\kappa_\alpha (t)  \Delta_{\alpha j} p_j(t) \label{eq:Ohm}
\end{equation}
where the conductance of edge $\alpha$ is
\begin{equation}
    \kappa_{\alpha}(t) = \frac{\pi r_{\alpha}(t)^4}{8 \mu l_{\alpha}(t)}.
\end{equation}

\begin{figure}
    \centering
    \includegraphics[width=0.7\linewidth]{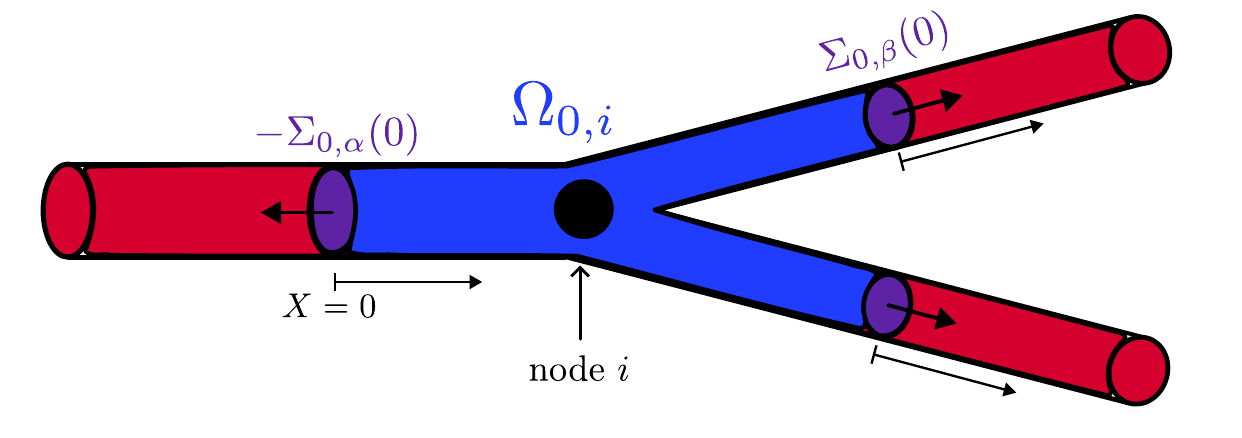}
    \caption{A typical junction in a network in the material configuration. To calculate the flow conservation law at node $i$, one can perform an integral of the continuity equation over the blue volume $\Omega_{0,i}$ which encloses half of each edge incident to node $i$. The arrows on the purple surfaces indicate the outward-pointing normal vector relative to the blue volume, while the arrows below the edges indicate the edge orientation. For the two edges on the right leaving node $i$ (indexed by $\beta$), these two vectors are aligned, but for the edge on the left entering node $i$ (indexed by $\alpha$), these two vectors are anti-aligned.  }
    \label{fig:KCL}
\end{figure}

Next, we need an equation analogous to Kirchhoff's Current Law which describes continuity of flow at the nodes. The easiest way to derive continuity at node $i$ is to integrate equation \eqref{eq:continuity_material_clean} over the volume $\Omega_{0,i}$ shown in figure \ref{fig:KCL}. The surface $\Sigma_{0,\alpha}(X)$ is the surface at fixed $X$ oriented in the positive $X$ direction along edge $\alpha$. The volume $\Omega_{0,i}$ is chosen so that each edge incident to node $i$ has half its rest volume inside $\Omega_{0,i}$, and each bounding surface lies at the center point of the tube. Integrating the left-hand side of \eqref{eq:continuity_material_clean} and applying the divergence theorem gives
\begin{align}
    \int_{\Omega_{0,i}} (\vec{\nabla} \cdot \vec{v}) \sqrt{|g|} \hspace{.1cm}dR \hspace{.1cm}  d\Phi \hspace{.1cm}  dX &= -\sum_{\alpha \rightarrow i} \int_{\Sigma_{0,\alpha}(0)} v^X  \sqrt{|g|} \hspace{.1cm}dR \hspace{.1cm}  d\Phi + \sum_{\beta \leftarrow i} \int_{\Sigma_{0,\beta}(0)} v^X \sqrt{|g|}\hspace{.1cm} dR \hspace{.1cm}  d\Phi \nonumber \\
    &= -\sum_{\alpha \rightarrow i} q_{\alpha}(0,t) + \sum_{\beta \leftarrow i} q_\beta(0,t) = -\sum_\alpha\Delta^T_{i \alpha} Q_\alpha(t).
\end{align}
The notation $\sum_{\alpha \rightarrow i}$ is understood to mean all edges entering node $i$ whereas $\sum_{\beta \rightarrow i}$ means all edges leaving node $i$. The minus sign on the $\sum_{\alpha \rightarrow i}$ term arises from a mismatch between the leftward-pointing normal vector and the rightward-pointing edge orientation. The other term in \eqref{eq:continuity_material_clean} integrates to
\begin{equation}
    \int_{\Omega_{0,i}} \left(\frac{1}{\sqrt{|g|}} \frac{\partial \sqrt{|g|}}{\partial t}\right) \sqrt{|g|} \hspace{.1cm}dR \hspace{.1cm}  d\Phi \hspace{.1cm}  dX = \frac{1}{2} \sum_{\alpha \rightarrow i} \frac{d \text{Vol}_\alpha}{d t} + \frac{1}{2} \sum_{\beta \leftarrow i} \frac{d \text{Vol}_\beta}{d t} = \frac{1}{2} \sum_\alpha |\Delta^T|_{i\alpha} \frac{d \text{Vol}_\alpha}{d t}. 
\end{equation}
Here, $|{\Delta}^T|_{i\alpha}$ is the unoriented incidence matrix obtained by taking the entry-wise absolute value of the incidence matrix, and 
\begin{equation}
    \text{Vol}_\alpha (t) = \pi r_\alpha(t)^2 l_\alpha (t).
\end{equation}
We are left with the following form of Kirchhoff's Current Law in our network:
\begin{equation}
    \sum_\alpha \Delta^T_{i\alpha} Q_\alpha(t) = \frac{1}{2} \sum_\alpha |\Delta^T|_{i\alpha }\frac{d \text{Vol}_\alpha(t)}{d t}  \label{eq:Kirchhoff}
\end{equation}
Intuitively, this equation states that any deviation from conservation of edge flow at a node is due to accumulation of volume in the blue region of figure \ref{fig:KCL}. 

Each edge is described by its current radius $r_\alpha(t)$ and length $l_\alpha(t)$ which parameterizes its conductance and volume. Combining equations \eqref{eq:Ohm} and \eqref{eq:Kirchhoff} gives us a single equation 
\begin{equation}
    \sum_{\alpha,j} \Delta^T_{i\alpha} \kappa_\alpha(t)  \Delta_{\alpha j} p_j(t) = -\frac{1}{2} \sum_{\alpha} |\Delta^T|_{i\alpha}\frac{d \text{Vol}_{\alpha}(t)}{d t}. 
\end{equation}
Once the pressure is specified at at least one location, the pressures at the remaining nodes can be obtained by inverting the matrix whose elements are $\sum_{\alpha}\Delta^T_{i\alpha} \kappa_\alpha(t)  \Delta_{\alpha j} $. Then, the edge flows can be calculated using \eqref{eq:Ohm}. 

The simple form of the network equations holds only for uniformly contracting edges. The simplification comes from the fact that, the derivation of Ohm's Law naturally involves the edge-averaged flow, while the derivation of Kirchhoff's Current Law naturally involves the flow at the midpoint of the tube; only for uniformly contracting edges are these the same. For more complicated contractions, one can still define an edge flow as either the edge-averaged flow or the flow at a particular point in an edge, but the network equations will involve additional terms to convert between the two. Furthermore, the flow on the edges will generically involve integrals like in equation \eqref{eq:q(X,t)_general} on each edge. In the next section, we will see that many uniformly contracting tubes in series can reproduce peristaltic waves, so computationally it may be advantageous to treat arbitrary contractions using the uniformly contracting edges.

As with the single-tube results, the Poisson ratio can be used to mechanically couple $r_\alpha$ and $l_\alpha$ once an additional assumption is made on the stress. We will only demonstrate how to solve the case of length-imposed peristalsis in networks. The radius-imposed case is more subtle since it may lead to geometrically incompatible length changes if equation \eqref{eq:ls_radial} is naively applied.

\subsection{Recovering longitudinal waves}
First, let us demonstrate how network models can be used to study length-imposed peristalsis. Consider a 1D linear network consisting of $N$ nodes and $N-1$ edges. Suppose the edge lengths satisfy 
\begin{equation}
    \frac{l_\alpha (t)-L_0}{L_0} = 1+ \epsilon \cos (2\pi( \alpha/N-t/T)). \label{eq:l_alpha_longitudinal}
\end{equation}
Note that the choice of edge lengths in a spatially embedded network is not arbitrary and must derive from node positions. It is possible that there are no node positions in an embedding space which will give rise to a set of edge lengths, or, as is the case here, there may be many such configurations. A simple choice of node positions for our purposes is one where the nodes all remain on the $x$-axis and node zero is fixed:
\begin{equation}
    x_i(t) = \begin{cases}
        0, & i=0\\
        \sum_{\alpha=1}^i l_\alpha (t), & i>0
    \end{cases} . \label{eq:positions_1D_network}
\end{equation}
The radii are obtained using equation \eqref{eq:ur_length_imposed}. Thus, all edge conductances $\kappa_{\alpha}$ and volume changes $\text{Vol}_\alpha$ can be calculated. Nodes $0$ and $N-1$ are kept at zero pressure and are open to allow for net in/outflow. The $T$-averaged edge flow $\langle Q_\alpha \rangle$ is the same for each $\alpha$, since no volume accumulates over one period. The flow is determined only by the lengths and not the positions. This is at the core of our choice of coordinate system. Because we have defined the flow with respect to the material coordinates of the tube wall, the flow does not depend on the Eulerian spatial coordinates. The entire network could even be translating or rotating, and it would not affect the result.

\begin{figure}
    \centering
    \includegraphics[width=\linewidth]{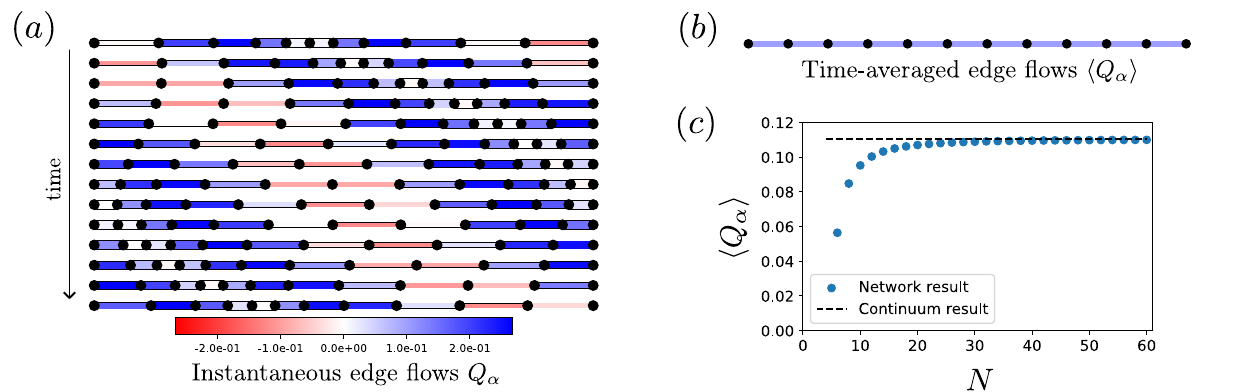}
    \caption{Demonstration of imposing node positions to study contraction-induced flows in a network. $(a)$ A 1D network consisting of $N=12$ nodes undergoes length-imposed peristalsis. Each row shows the current network geometry $\Omega$ and flow at an instant in time, and the full period is plotted in 14 frames. The edge color is the instantaneous flow through the edge with all edges oriented from left to right, so red indicates leftward flow and blue indicates rightward flow. The length and width of the edges are proportional to $l_\alpha$ and $r_\alpha$, respectively. $(b)$ The time-average flow is plotted for each edge in the tube's rest frame $\Omega_0$. All edges have the same average flow. $(c)$ The time-average flow is plotted as a function of $N$. For large $N$, the network model approaches the result for a continuous peristaltic wave. The wavelength was chosen to be the system size so as to recover periodic boundary conditions. In all cases, the boundary nodes have a fixed position $p=0$, and the positions of the nodes are given by \eqref{eq:positions_1D_network}, with $\epsilon = 0.5$ and $\nu = 0.5$. }
    \label{fig:network1D}
\end{figure}

An example network and resulting flow is plotted in figure \ref{fig:network1D} with $N=12$, $\epsilon = 0.5$, $\nu = 0.5$, and the boundary nodes fixed to zero pressure. The network is driven by the displacement of the nodes which induces length changes dictated by equation \eqref{eq:l_alpha_longitudinal}, and radius changes calculated from equation \eqref{eq:ur_length_imposed}, and thus changes in edge volume and conductance. The instantaneous flows in panel $(a)$ are different for each edge, with positive flows (from left to right) labeled in blue and negative flows (from right to left) labeled in red. The time-averaged flow is conveniently plotted in the material frame $\Omega_0$ in panel $(b)$. Each edge has identical mean flow. As the density of nodes increases, the network approaches the continuous peristaltic wave result given by equation \eqref{eq:general_mean_flow}, and displayed with a dashed line in panel $(c)$.

With only slightly more effort, all of the continuum results considered so far can be recovered, and more. A nonzero pressure difference can be applied across the endpoints to incorporate pressure effects. Obviously, if there is only one edge, this recovers the results for a uniformly contracting tube. In the regimes where radius-imposed peristalsis is valid, one can calculate the edge lengths using \eqref{eq:dudx_radius_imposed} and the positions using \eqref{eq:positions_1D_network}, and simulate discrete radial peristaltic waves. In general, for a 1D network of edges in series, one can choose any positive-valued functions $r_\alpha (t)$ and $l_\alpha (t)$ to simulate contraction-induced flow in a tube. For example, it was shown that there is an optimal phase lag for 1D channels of radially contracting vessels \cite{Amselem2023ValvelessNumbers}.

\subsection{Contraction by imposed node positions}
The method in the previous subsection was developed to reproduce the continuum theory. Instead of defining the edge lengths, we could have defined the time-dependent node positions. If node positions are known, then the strain $\epsilon_{XX}$ can be calculated, and if an additional assumption is made on the stress of each edge (for example, $\sigma_{\Phi \Phi} = 0$), then $\epsilon_{\Phi \Phi}$ and therefore the radii can be calculated. Because the node positions are given at the outset, there is no concern regarding geometric incompatibility, and there is no challenge incorporating nonlinear strains as we have done throughout the paper.

Note that it is always possible to go from a change in node positions to a change in length to a change in radius, but if we instead attempt to impose a radius, there is no guarantee that the calculated change in length using equation \eqref{eq:ls_radial} will correspond to a physical change in node positions. Thus, radius-imposed peristalsis is more subtle than length-imposed peristalsis in networks. Although, if the length is held fixed, this is not an issue, as is done when studying the slime mold \cite{Alim2013RandomIndividual}.

\section{Discussion}

We studied the role of longitudinal effects in persitaltic pumping by considering two different boundary conditions: radius-imposed peristalsis where the tube is free to deform longitudinally, and length-imposed peristalsis where the tube is free to deform radially. Employing the membrane theory for a thin, axisymmetric tube, free boundary conditions were used to couple the two components of the in-plane strain, and thus couple radial wall motion to longitudinal wall motion through the Poisson ratio $\nu$.

When studying radius-imposed peristalsis, we found several generalizations to the fixed-length ($\nu = 0$) flow calculations. Materials with $\nu > 0$ have suppressed flow since a large radius necessarily produces a small length, thus reducing the magnitude of volume changes. On the other hand, materials with $\nu < 0$ have enhanced flow. This suggests that peristaltic pumps engineered with auxetic materials may improve pumping performance. Walker and Shelley calculated optimal shapes for peristaltic waves in two-dimensional regions with inextensible boundaries \cite{Walker2010ShapePumping}. Our work suggests that allowing for extensible boundaries could further improve performance, though their work optimized for power which we did not consider. 

The Poisson ratio also affects particle trajectories. In an incompressible tube driven by radius-imposed peristalsis, both trapping and reflux are suppressed as compared to a tube of fixed length. For small enough adverse pressures, no reflux exists for $\nu >0$, unlike the result for a tube of fixed length where all pressures generate reflux \cite{Shapiro_Jaffrin_Weinberg_1969}. It has been suggested that reflux in the ureter could transport bacteria from the bladder to the kidneys \cite{Kalayeh_Xie_Brian_Fowlkes_Sack_Schultz_2023, Shapiro_Jaffrin_Weinberg_1969}. Our findings suggest that if there were no longitudinal wall motion in the ureter, then reflux would be even more prevalent. Kalayeh, et al. found that reflux is suppressed when the longitudinal wall velocity is in the wave direction during expansion and against the wave direction during contraction \cite{Kalayeh_Xie_Brian_Fowlkes_Sack_Schultz_2023}. Their analysis considered Eulerian velocity boundary conditions where the longitudinal and transverse wall velocities were assigned arbitrarily. Remarkably, an incompressible solid naturally produces this coupling between expansion and longitudinal wall velocity (see figure \ref{fig:radialVelocityField}$(d)$, for example), so even though our approach incorporated more accurate modeling of the tube wall, our findings concerning reflux agree with those of Kalayeh, et al. and further explain the observations of longitudinal wall motion in ureteral peristalsis.

For a material undergoing length-imposed peristalsis, backflow is not a special property close to the wall, but is actually typical for all streamlines, provided the amplitude is not too large. Flow opposite the wave direction is a surprising phenomenon that has also been documented in models for peristaltic pumping of non-Newtonian fluids \cite{Provost1994APumping} and peristaltic pumping with valves \cite{Winn2024OperatingValves}. It was also found in Kalayeh, et al. that large-amplitude longitudinal waves require large-amplitude radial waves in order to obtain positive flow \cite{Kalayeh_Xie_Brian_Fowlkes_Sack_Schultz_2023}. Local segments of the esophagus are reported to shorten by as much as 0.34 times their original length when undergoing longitudinal muscle contractions \cite{Nicosia2001LocalUltrasonography}. This is far beyond the regime where linear elasticity applies, so our results concerning nonlinear strain could be relevant in the esophagus. The magnitude of flow for length-imposed peristalsis is sensitive to the Poisson ratio, largely due to the factor of $r_s^4$ in the conductance. Most notably, the flow terms quadratic in the strain vanish when $\nu = 1/2$, indicating that small-amplitude length-imposed peristalsis produces very little flow, which limits its usefulness as a mechanism for flow propulsion. However, we found that at large enough amplitude, in an incompressible tube, trapping may still occur leading to large flow in the direction of the traveling wave. This is all relevant for the esophagus which is known to be incompressible \cite{Nicosia2001LocalUltrasonography}. In a computational study of the esophagus, it was argued that longitudinal muscle alone could not transport a bolus \cite{KouEtAl2015}. While we agree that it is much more difficult to trap a bolus with longitudinal muscle as compared with radial muscle, it should be possible if the amplitude of contraction is large enough.

The free boundary conditions used in this paper produce a simple coupling between the two in-plane strains, but this is only one of many reasonable boundary conditions to consider. It is worth mentioning that the typical study of radius-imposed peristalsis with fixed length could be valid even for a material with $\nu \neq 0$ provided that the material is confined in such a way that longitudinal motion is suppressed. There are even cases in which a material with positive Poisson ratio can have in-phase radial and longitudinal contractions (effectively acting like a free material with negative Poisson ratio). For example, if the boundary deformation is driven only by the pulsatile flow inside a compliant vessel (such as in the arteries), the radius and length are in phase, though the longitudinal displacement vanishes for $\nu = 1/2$ \cite{Canic2003EffectiveArteries}. This is a qualitative difference between pulsatile flow and peristaltic flow. For the former, the components of the stress are coupled together through the fluid: pressure exerts normal stress while pressure gradients exert shear stress. For the latter, the stresses are determined by the muscle structure generating tension in the tube. In our model, we assumed that one component of the in-plane stress is zero and that fluid-to-solid interaction is negligible. A more general formulation would require using Laplace's law \cite{Tang1993NumericalBoundaries}: A balance between external peristaltic forces and fluid forces determines how the two components of the in-plane strain are coupled \cite{Takagi_Balmforth_2011, Carew_Pedley_1997, Elbaz_Gat_2014}. This would introduce explicit dependence on the Young's modulus. Softness tends to cause a phase shift in stress and strain, while also suppressing flow at a fixed strain amplitude for radius-imposed waves \cite{Winn2024OperatingValves}. It was recently demonstrated that softness may increase flow when driven by longitudinal waves, an effect which could be relevant for studying perivascular pumping in the brain \cite{Trevino2025LowSpace}. Only when the fluid pressure is much smaller than the stiffness of the vessel can the softness effects be neglected. Additional aspects of the mechanics neglected in this paper include the separate role of active versus passive wall tensions and the possibility of a non-Hookean elastic response, both of which are relevant for describing contractions by smooth muscle \cite{Carew_Pedley_1997,Fung1993}, though our paper focused only on the passive coupling between nonlinear strain components.

For networks of radially contracting vessels with $\nu \neq 0$, incorporating some additional mechanics information is not only more accurate, but is also necessary to ensure that the lengths of the edges correspond to differences in node positions. This is not an issue for networks where the node positions are given as input to calculate the flow. We showed that the flow induced by oscillating nodes in a dense 1D network converges to the continuous length-imposed peristalsis result. This technique can be generalized to complex networks. For example, it has been hypothesized that the gastrovascular system of the moon jellyfish \textit{Aurelia aurita}, a network of branching canals, is partially driven by full-body contractions which stretch the vessels, in addition to cilia \cite{Southward1955ObservationsL.}. By tracking the motion of a swimming jellyfish, one could use our model to estimate the internal flows that would be produced by whole body deformations. This is only possible because the flow, as we have defined it, is measured with respect to material coordinates which travel with the swimming jellyfish, as opposed to Eulerian coordinates which are fixed in space.

In order to solve the problem of contraction-induced flows in tubes undergoing large strain, we analyzed the fluid in a fixed domain using the Lagrangian coordinates of the solid and a time-dependent metric. This is a novel approach to solving peristalsis problems. This method has four advantages: It leads to a clear definition of the flow with respect to the moving boundary \eqref{eq:q_definition}; it simplifies boundary conditions and makes it easier to enforce periodicity; it makes it possible to incorporate finite strain using a Lagrangian description of elasticity; and it is easily generalized to networks. In addition to allowing for explicit solutions, this method provides new intuition for the problem. The velocity field can be understood as the sum of a parabolic profile in the material configuration scaled by the local stretch and a boundary term. Only the parabolic term is relevant for calculating the flow. The metric acts as an effective source term generating a divergence in an otherwise incompressible fluid, as described by equation \eqref{eq:continuity_material}. This geometric approach to contraction-induced flows could be used to study other modes of deformation. Relaxing the axisymmetric assumption would allow for the study of fluid flows induced by bending modes. The centerline of the rest configuration need not even be a line. The geometric approach used in this paper only assumed that a flat surface $\Sigma_0(X)$ in the material configuration bounded by the undeformed tube is mapped to a new surface $\Sigma(X,t)$ in the current configuration bounded by the deformed tube, where $X$ is in general a coordinate parameterizing the centerline. Laminar flow has been studied in some idealized curved stationary tube geometries \cite{Murata1976LaminarCurvature, Wang1981OnPipe}, and our method provides a way to generalize these calculations to dynamic tube geometries. 

Our geometric method of studying a contraction-induced flow is closely related to the Arbitrary Lagrangian-Eulerian (ALE) method \cite{Donea2004Methods}. 
In the ALE method, a structure is studied computationally using a mesh whose movement may follow the particles (Lagrangian), be fixed in space (Eulerian), or have any other arbitrary motion. The mesh motion may be imposed by an interpolation to known displacements at the system boundary or solved for using a mesh adaptation algorithm. The ALE method is appreciated for its applications to problems with deforming boundaries and interfaces, though the mesh motion is typically not thought of as a physical quantity. The moving mesh is typically chosen to prevent large distortions and carefully partition structures at an interface to ensure the computation converges to the correct solution. Instead, here we are claiming that this moving mesh has some physical significance. Since all quantities should be measured with respect to the moving boundary, a reasonably chosen mesh gives us a good indicator of the physical movement with respect to the boundary. Note that our mesh motion is not completely arbitrary, but is also not unique. We require that the mesh becomes Lagrangian at the fluid-solid interface, but how exactly we interpolate the mesh in the fluid region is less obvious. For problems of flow through deforming tubes, the choice of equation \eqref{eq:xSigma} is the simplest and most intuitive to describe the velocity field. Although the method presented in this paper was developed for analytical purposes, it was also implemented as an ALE mesh in COMSOL simulations (see Appendix).

The material coordinates used to analyze peristalsis could be applied to a broad array of problems in fluid dynamics where boundary deformations drive flow. For example, the peristalsis problem closely resembles earlier work by Taylor describing the fluid flow near a single inextensible undulating boundary used to model the flagella of a swimming micro-organism \cite{Taylor_1951}. The swimming speed is calculated by assuming the velocity decays at infinity and the net force on the swimmer is zero. It was later pointed out that if instead a zero displacement constraint were imposed, the undulating boundary would act like a pump: the force exerted by the boundary to prevent net displacement would drive fluid flow in the direction of the traveling wave \cite{Lauga2009TheMicroorganisms}. The swimming direction is opposite the direction of the propagating wave, much like the flows generated by length-imposed peristaltic waves. Thus, as the problems of determining the flow generated by an undulating boundary and determining the swimming speed of an undulating boundary are closely related, we believe our method of studying fluid flows using the Lagrangian coordinates of the elastic boundary could also be applied to the study of swimming organisms.

\section{Appendix}
\subsection{COMSOL Simulations}
\begin{figure}
    \centering
    \includegraphics[width=0.5\linewidth]{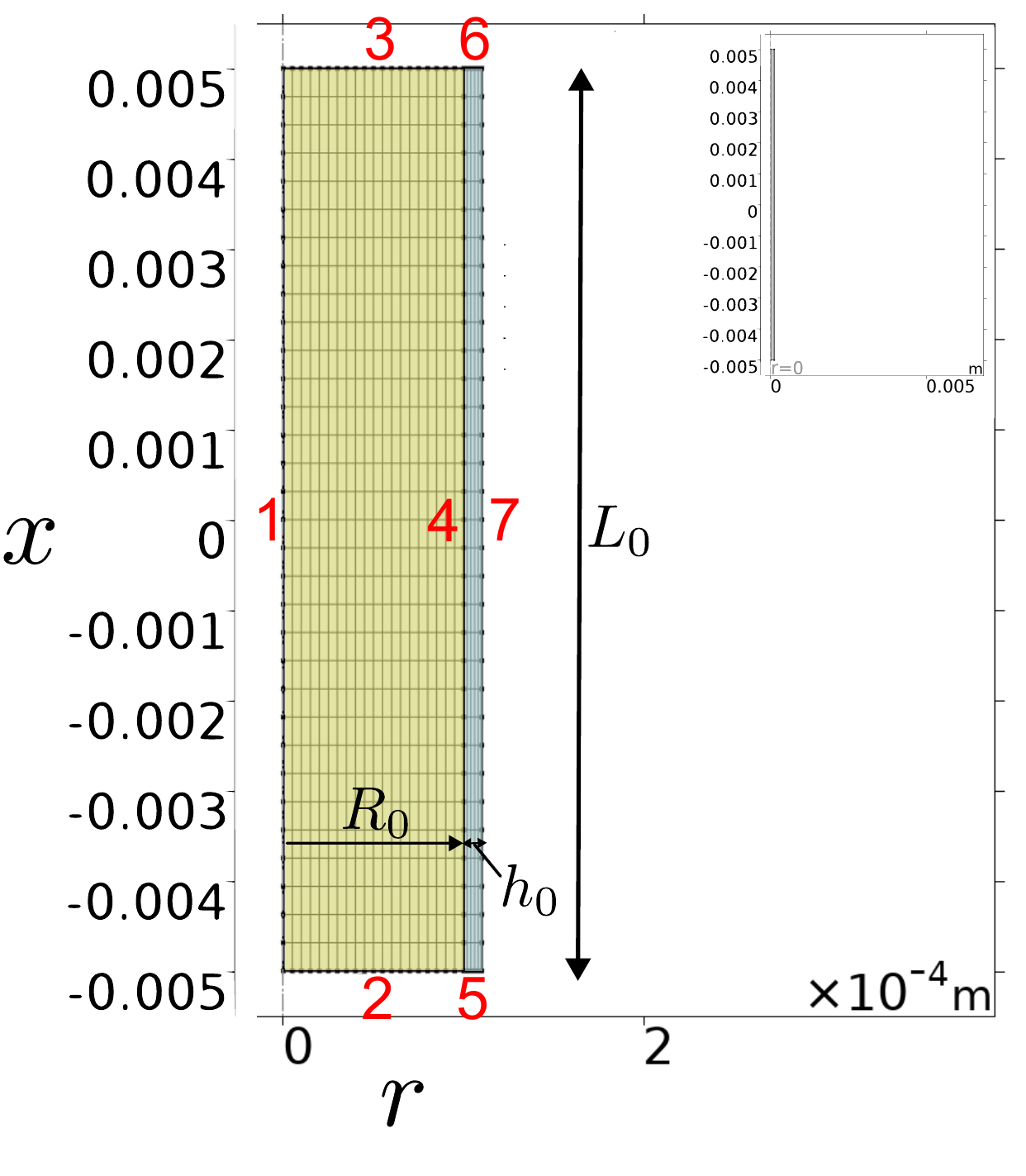}
    \caption{2D axisymmetric geometry used for COMSOL simulations. The boundaries are labeled with red numbers for reference. The solid is colored blue, and the fluid is colored yellow. An example mesh is displayed. The radius is scaled by a factor of 20 for visualization purposes. The actual relative dimensions are displayed in the inset in the top right.}
    \label{fig:COMSOL}
\end{figure}

In order to test the predictions of the model, analytic results were compared to finite element simulations in COMSOL 6.3. All simulations were performed in a 2D axisymmetric geometry. The tube at rest is identical to that in figure \ref{fig:CoordinateSystems}, except that a finite thickness $h_0$ is used. The fluid obeys the Navier Stokes equations for an incompressible, Newtonian fluid, while the solid obeys the Cauchy momentum equation and Hooke's law, keeping the geometric nonlinearity. The fluid and solid stress and velocity fields are fully coupled at their interface using the fluid-structure interaction module, though for some analyses, the tube's Young's modulus is set to a large value to suppress fluid to solid coupling. No adaptive meshing is used, and the mesh motion in the fluid domain is instead prescribed to follow the solid's motion, analogous to equation \eqref{eq:xSigma}. This prescribed mesh deformation is achieved by using the general extrusion operator from the inner tube displacement field to the bulk fluid and initializing with a rectangular mesh. Figure \ref{fig:COMSOL} displays the geometry and mesh with boundaries labeled for reference. Boundary 1 is the symmetry line $r=0$, while the remaining boundary conditions will vary depending on the study. The initial conditions for all quantities are set equal to zero, and time-dependent simulations gradually ramp up any boundary conditions such that the boundary condition and its first three derivatives all equal zero at $t=0$. We will continue to use $x$ to refer to the axial coordinate, though in COMSOL, the axial coordinate is denoted $z$.

\subsubsection{Uniform radius-imposed contractions}
For the simulations of uniform radial contraction, $u_r(X,t)$ is imposed on boundary 7 and $u_x(R,t)$ is set equal to zero on boundary 5 to ensure no global motion. Free boundary conditions are used throughout the rest of the solid. At boundary 2, a normal stress of $\Delta p * t^4/(.1+t^4)$ is applied. At boundary 3, the normal stress is fixed to zero. 

In order to generate the plots shown in figure \ref{fig:UniformContraction}, a time-dependent simulation sweeps through all desired values of $u_r$ quasi-statically. Note that although the radius is imposed on the outer boundary, it is the inner boundary which is relevant for driving fluid flows. The following information is recorded as a function of time: $u_r$ averaged over boundary 4, $u_x(R_0, L_0/2)$, the integral of 1 over the current deformed fluid domain, and the ratio of the flow through 2 and the pressure drop $\Delta p$. These quantities are used to calculate $r_s$, $l_s$, $\text{Vol}$, and $\kappa$. Note that because boundary 2 is fixed (since boundary 5 is fixed in the $x$-direction), there is no ambiguity in how to define the flow, and although the flow contains both pressure and contraction terms, the contraction terms are negligible at the expansion rate and pressure drop used in this simulation. 

The initial mesh used for these simulations was constructed using 33 equally-spaced lines of constant $x$, 21 equally-spaced lines of constant $r$ in the fluid region, and 6 lines of constant $r$ in the solid region. There are a total of $32 \times 20 = 640$ rectangular elements of equal size in the fluid domain and $32 \times 5 = 160$ rectangular elements of equal size in the solid domain. This is exactly the mesh shown in figure \ref{fig:COMSOL}. The mesh in the solid domain is Lagrangian: it follows the displacement field of the solid, while the the mesh in the fluid domain obeys equation \eqref{eq:xSigma}. Although the form of \eqref{eq:xSigma} was chosen for convenience in analytic calculations, it is interesting to note that these mesh dynamics are useful in finite element simulations even with rather large elements.

The physical parameters used are inspired by microvasculature: $L_0 = 10^{-2}$ m, $R_0 = 10^{-4}$ m, $h_0 = 10^{-5}$ m, $\Delta p = 100$ Pa, Young's modulus $E = 10^6$ Pa, $\mu = 10^{-3}$ Pa$\cdot$s, fluid density $\rho = 1000$ kg/m$^3$, and solid density $\rho_{\text{tube}} = 1025$ kg/m$^3$. This set of parameters ensures the fluid is at low Reynolds number and the lubrication theory holds. As long as the approximations used to derive equations \eqref{eq:momentum_integrated} hold, then the normalized quantities in figure \ref{fig:UniformContraction} are independent of all of these parameters. A parameter sweep was performed on the Poisson ratio $\nu$ to generate the data in figure \ref{fig:UniformContraction}. Note that COMSOL does not allow for a Poisson ratio of $\nu = 0.5$, so an almost incompressible solid of $\nu=0.49$ was used instead. 

In order for the time-dependent simulation to represent a quasistatic deformation (with negligible fluid-to-solid interaction), the deformation must occur sufficiently slow. The imposed radial displacement was chosen to ensure this:
\begin{equation}
    u_r(t) = \frac{t^4}{1+t^4}\left[\min(u_r) + \left(\max(u_r)-\min(u_r)\right)\text{Ramp}\left(\frac{t-10}{200}\right) \right], \label{eq:ramp}
\end{equation}
where the Ramp$(x)$ is equal to zero for $x<0$ and equal to $x$ for $x>0$, and $\min(u_r)$ and $\max(u_r)$ are the specified minimum and maximum displacements. The pre-factor ensures the initial conditions are satisfied. The function falls to $\min(u_r) = -0.8R_0$, and at $t=10$ s, rises linearly to $\max(u_r) = 0.65 R_0$ over a duration of 200 s. The lower magnitude of $\max(u_r)$ as compared to $\min(u_r)$ was chosen to ensure that the tube length remains positive for $\nu = 0.49$, according to equation \eqref{eq:max_amplitude}.

Although $\sigma_{XX}=0$ is only enforced at boundaries 5 and 6, along with $\sigma_{RX}=0$ on boundary 7, it is observed that $\sigma_{XX}$ remains at least an order of magnitude smaller than $\sigma_{\Phi \Phi}$ throughout the simulation, so \eqref{eq:dudx_radius_imposed} holds approximately. The agreement between analytic results and COMSOL, suggests our analytic results should hold for some realistic choices of parameters. 

\subsubsection{Uniform length-imposed contractions}
For the simulations of uniform longitudinal contraction, $u_x(R,t)$ is imposed on boundaries 6 and $u_x(R,t)$ is set equal to zero on boundary 5 to ensure no global motion. Furthermore, $u_x(X,t)=u_x(L_0/2,t)*(X+L_0/2)/L_0$ is imposed on boundary 7. The functional form for the imposed axial displacement is exactly equation \eqref{eq:ramp}, only $u_r$ is replaced with $u_x$. 

The mesh and physical parameters used are the same as that for the uniform radius-imposed contractions. 

The remaining boundary conditions are free: $\sigma_{RR}=0$ on boundary 7, $\sigma_{RX}=0$ on boundaries 5 and 6. Although nowhere is it enforced that $\sigma_{\Phi \Phi} = 0$, $\sigma_{XX}$ is always orders of magnitude bigger than the other components, so equation \eqref{eq:ur_length_imposed} holds approximately, and we again get good agreement between analytics and COMSOL as shown in figure \ref{fig:UniformContraction}.

\subsubsection{Radius-imposed peristaltic waves}

For the simulations of radius-imposed peristaltic waves, the following boundary conditions were imposed on boundary 7:
\begin{align}
    u_r(X,t) &= \epsilon R_0 \left(\frac{t^4}{.3+t^4} \right) \sin (2\pi(X/\lambda - t/T)) \\
    \sigma_{RX}(X,t) &= 0.
\end{align}
One wavelength was simulated with appropriate periodicity conditions. A periodic flow condition was applied on boundaries 2 and 3 with pressure difference $\Delta p = (8c\mu \lambda/R_0^2 ) \Delta \bar{p}_\lambda$. An additional constraint needs to be given on the pressure, but fixing the pressure at a single point caused the simulations to fail. Instead, an integral constraint was imposed, keeping the spatially average pressure fixed to zero. On boundary 5, $u_x$ was fixed to zero to prevent rigid-body motion. A periodic condition enforced that $u_r$ was identical on boundaries 5 and 6, but $u_x$ was kept free on boundary 6 to allow for global shortening. 

In order to generate the points in figure \ref{fig:flow_radial_wave}$(d)$, a parameter sweep was performed on the different values of $\epsilon$ and $\nu$ at a fixed $\Delta \bar{p}_{\lambda} = 0$. For each simulation in the parameter sweep, a time-dependent study was performed until $t=4$ (four periods). This was sufficient to get convergence to the steady state solution. The flow through boundary 2 is calculated. There is no ambiguity in defining the flow since the boundaries 2 and 5 are stationary. The same procedure was performed to obtain figure \ref{fig:flow_radial_wave}$(h)$, except that a value of $\Delta \bar{p}_{\lambda} = 1$ was used. 

A rectangular mesh was again used, generated using 25 equally-spaced lines of constant $x$, 25 equally-spaced lines of constant $r$ in the fluid region, and 6 lines of constant $r$ in the solid region. There are a total of $24 \times 24 = 576$ rectangular elements of equal size in the fluid domain and $24 \times 5 = 120$ rectangular elements of equal size in the solid domain. 

The physical parameters were slightly altered from the uniform contractions case to ensure that the assumptions on the forces were still satisfied. The thickness was reduced to $h_0 = 2\times 10^{-6}$ m to ensure the membrane theory applies, and the Young's modulus was increased to $E = 0.5 \times 10^9$ Pa to prevent fluid-to-solid interaction. The wavelength was the same as the tube length for the uniform contractions: $\lambda = 10^{-2}$ m. The $\epsilon$ values used for the parameter sweep were $\{0.05, 0.20, 0.35, 0.50, 0.65 \}$. All other parameters were the same as for the uniform contractions. 

Note that nowhere is the in-plane strain $\sigma_{XX}$ imposed, yet it remains much smaller than $\sigma_{\Phi \Phi}$, so \eqref{eq:dudx_radius_imposed} holds approximately.

\subsubsection{Length-imposed peristaltic waves}
For the simulations of length-imposed peristaltic waves, the following boundary conditions were imposed on boundary 7:
\begin{align}
    u_x(X,t) &= \frac{\epsilon \lambda}{2\pi} \left(\frac{t^4}{.3+t^4} \right) \left[ \sin (2\pi(X/\lambda - t/T)) - \sin (2\pi(-L/2/\lambda - t/T))\right]\\
    \sigma_{RR}(X,t)  &= 0.
\end{align}
The second sine function ensures that the boundary at $-L/2$ remains fixed, so that the flow can be calculated unambiguously through boundary 2. The periodicity conditions and parameters are the same as that for the radius-imposed waves.

\subsection{Flow calculations}
In this appendix, we derive the equation for the flow in a tube undergoing arbitrary contractions, equation \eqref{eq:q(X,t)_general} in the main text. From this, we will derive the flow for a uniformly contracting tube and a tube undergoing peristaltic contractions, equations \eqref{eq:qUniform} and \eqref{eq:general_mean_flow} in the main text.

We begin with equations \eqref{eq:continuity_integrated} and \eqref{eq:momentum_integrated} and a known pressure drop between points $X_1$ and $X_2$. First, integrate equation \eqref{eq:continuity_integrated} from $X_1$ to $X$:
\begin{equation}
    q(X,t) = q(X_1,t) - \int_{X_1}^{X} \frac{\partial }{\partial t} \left( \pi r_s(X'')^2 \frac{\partial x_s(X'')}{\partial X} \right) dX''. \label{eq:q(X,t)_X1}
\end{equation}
Next, integrate equation \eqref{eq:momentum_integrated} from $X_1$ to $X_2$ and apply the definition $\Delta p(t) \equiv p(X_2,t) - p(X_1,t)$:
\begin{equation}
    \Delta p(t) + \int_{X_1}^{X_2} \frac{8\mu}{\pi r_s(X')^4} \frac{\partial x_s (X')}{\partial X} \left[q(X_1,t) - \int_{X_1}^{X'} \frac{\partial }{\partial t} \left( \pi r_s(X'')^2 \frac{\partial x_s(X'')}{\partial X} \right) dX'' \right]dX' = 0,
\end{equation}
\begin{equation}
    \implies q(X_1,t) = \frac{-\Delta p(t)}{\int_{X_1}^{X_2} \frac{8\mu}{\pi r_s(X')^4} \frac{\partial x_s(X')}{\partial X} dX'} + \frac{\int_{X_1}^{X_2} \frac{8\mu }{\pi r_s(X')^4} \frac{\partial x_s(X')}{\partial X} \left(\int_{X_1}^{X'} \frac{\partial}{\partial t}\left(\pi r_s(X'')^2 \frac{\partial x_s(X'')}{\partial X} \right) dX'' \right)dX'}{\int_{X_1}^{X_2} \frac{8\mu}{\pi r_s(X')^4} \frac{\partial x_s(X')}{\partial X}  dX'}.
\end{equation}
Plugging back into equation \eqref{eq:q(X,t)_X1}, 
\begin{align}
    q(X,t)&= \frac{-\Delta p(t)}{\int_{X_1}^{X_2} \frac{8\mu }{\pi r_s(X')^4} \frac{\partial x_s(X')}{\partial X} dX'} - \frac{\int_{X_1}^{X_2} \frac{8\mu }{\pi r_s(X')^4} \frac{\partial x_s(X')}{\partial X} \left(\int_{X_1}^{X'} \frac{\partial}{\partial t}\left(\pi r_s(X'')^2 \frac{\partial x_s(X'')}{\partial X} \right) dX'' \right)dX'}{\int_{X_1}^{X_2} \frac{8\mu}{\pi r_s(X')^4} \frac{\partial x_s(X')}{\partial X}  dX'} \nonumber\\
    &\hspace{.4cm}-\int_{X_1}^{X} \frac{\partial }{\partial t} \left( \pi r_s(X'')^2 \frac{\partial x_s(X'')}{\partial X} \right) dX'. 
\end{align}
Expressing these terms with a common denominator and simplifying the resulting expression produces equation \eqref{eq:q(X,t)_general}, which is repeated here for convenience:
\begin{align}
    q(X,t)&= \frac{-\Delta p(t)}{\int_{X_1}^{X_2} \frac{8\mu }{\pi r_s(X')^4} \frac{\partial x_s(X')}{\partial X} dX'} - \frac{\int_{X_1}^{X_2} \frac{8\mu }{\pi r_s(X')^4} \frac{\partial x_s(X')}{\partial X} \left(\int_{X}^{X'} \frac{\partial}{\partial t}\left(\pi r_s(X'')^2 \frac{\partial x_s(X'')}{\partial X} \right) dX'' \right)dX'}{\int_{X_1}^{X_2} \frac{8\mu}{\pi r_s(X')^4} \frac{\partial x_s(X')}{\partial X}  dX'}. \label{eq:q(X,t)_general_again}
\end{align}
Writing $q(X,t)$ in this way stresses the interaction between two mechanisms involved in peristalsis: viscous resistance and volume change. 

For the case of uniform contractions, $r_s$ and $\frac{\partial x_s}{\partial X}$ are $X$-independent, so one can pull all of the geometric factors out of the integrals in \eqref{eq:q(X,t)_general_again}, leaving only a single factor of $X$ in $q(X,t)$ arising from the upper bound in the second term. This gives equation \eqref{eq:qUniform}.

For the case of waves, one can make the replacement $\frac{\partial}{\partial t} \rightarrow -c \frac{\partial}{\partial X}$ in the inner integral in the second term of equation \eqref{eq:q(X,t)_general_again}, then apply the fundamental theorem of calculus. This leads to 
\begin{align}
    q(X,t)&= \frac{-\Delta p(t)}{\int_{X_1}^{X_2} \frac{8\mu }{\pi r_s(X')^4} \frac{\partial x_s(X')}{\partial X} dX'} - \frac{\int_{X_1}^{X_2} \frac{8\mu }{\pi r_s(X')^4} \frac{\partial x_s(X')}{\partial X} \left(c\pi r_s(X')^2 \frac{\partial x_s(X')}{\partial X} - c\pi r_s(X)^2 \frac{\partial x_s(X)}{\partial X} \right)dX'}{\int_{X_1}^{X_2} \frac{8\mu}{\pi r_s(X')^4} \frac{\partial x_s(X')}{\partial X}  dX'} \nonumber \\
    &= c\pi R_0^2 \left[ \left(\frac{r_s(X)}{R_0}\right)^2 \left(\frac{\partial x_s(X)}{\partial X}\right) + \frac{-\Delta p(t)/(8c\mu/R_0^2)}{\int_{X_1}^{X_2} \left(\frac{ r_s(X')}{R_0}\right)^{-4} \left( \frac{\partial x_s(X')}{\partial X}\right) dX'}  - \frac{\int_{X_1}^{X_2}\left(\frac{r_s(X')}{R_0}\right)^{-2} \left(\frac{\partial x_s(X')}{\partial X}\right)^2 dX'}{\int_{X_1}^{X_2} \left(\frac{r_s(X')}{R_0} \right)^{-4} \left(\frac{\partial x_s(X')}{\partial X} \right) dX'}\right].
\end{align}
So far, the solution is valid even for a finite tube where flow is unsteady, generalizing the result in \cite{Li1993Non-SteadyTubes}. The spatial integrals can, in general, depend on time. In the special case of an infinitely long tube, we can take $X_1 = 0$ and $X_2 = \lambda$, and the spatial integrals over one wavelength are time-independent and proportional to time integrals over one period. The flow in an infinitely long tube, using $\langle \cdot \rangle$ to denote $T$-averages, is
\begin{equation}
     q (X,t) = c\pi R_0^2 \left[\Big(\frac{r_s(X)}{R_0}\Big)^2 \Big(\frac{\partial x_s(X)}{\partial X}\Big)  -\frac{ \Delta \bar{p}_{\lambda}(t)}{\langle (\frac{r_s}{R_0})^{-4} \big(\frac{\partial x_s}{\partial X} \big) \rangle }  -  \frac{\langle (\frac{r_s}{R_0})^{-2} \big(\frac{\partial x_s}{\partial X}\big)^2 \rangle }{\langle (\frac{r_s}{R_0})^{-4} \big(\frac{\partial x_s}{\partial X}\big) \rangle } \right],
\end{equation}
where $\Delta \bar{p}_{\lambda} \equiv [p(\lambda)-p(0)]/(8c \mu \lambda / R_0^2)$. Only the first term depends on space and time, and only in the combination $X-ct$. Taking the time-average of this equation recovers \eqref{eq:general_mean_flow}.

\begin{acknowledgments}
This research was supported by the University of Pennsylvania Materials Research Science and Engineering Center (MRSEC) through Award DMR-2309043, the HFSP Award No. RGP015/2023 and the John Templeton Foundation through the support of Grant 62846.
    
\end{acknowledgments}

\bibliographystyle{unsrt}
\bibliography{refs}

\begin{thebibliography}{10}

\bibitem{Nicosia2001LocalUltrasonography}
Mark~A Nicosia, James~G Brasseur, Ji-Bin Liu, and Larry~S Miller.
\newblock {Local longitudinal muscle shortening of the human esophagus from high-frequency ultrasonography}.
\newblock {\em Am J Physiol Gastrointest Liver Physiol}, 281:1022--1033, October 2001.

\bibitem{Pal2002TheTransport}
Anupam Pal and James~G. Brasseur.
\newblock {The mechanical advantage of local longitudinal shortening on peristaltic transport}.
\newblock {\em Journal of Biomechanical Engineering}, 124(1):94--100, 2002.

\bibitem{BrasseurEtAl2007}
James~G. Brasseur, Mark~A. Nicosia, Anupam Pal, and Larry~S. Miller.
\newblock Function of longitudinal vs circular muscle fibers in esophageal peristalsis, deduced with mathematical modeling.
\newblock {\em World Journal of Gastroenterology}, 13(9):1335--1346, 2007.

\bibitem{Carew_Pedley_1997}
E.~O. Carew and T.~J. Pedley.
\newblock An active membrane model for peristaltic pumping: Part i—periodic activation waves in an infinite tube.
\newblock {\em Journal of Biomechanical Engineering}, 119(1):66–76, February 1997.

\bibitem{Kalayeh_Xie_Brian_Fowlkes_Sack_Schultz_2023}
Kourosh Kalayeh, Haotian Xie, J.~Brian~Fowlkes, Bryan~S. Sack, and William~W. Schultz.
\newblock Longitudinal wall motion during peristalsis and its effect on reflux.
\newblock {\em Journal of Fluid Mechanics}, 964:A30, 2023.

\bibitem{Moore2018LymphaticFlows}
James~E. Moore and Christopher~D. Bertram.
\newblock {Lymphatic System Flows}.
\newblock {\em Annual Review of Fluid Mechanics}, 50:459--482, January 2018.

\bibitem{Wolf2021FluidValves}
Ki~Tae Wolf, J.~Brandon Dixon, and Alexander Alexeev.
\newblock {Fluid pumping of peristaltic vessel fitted with elastic valves}.
\newblock {\em Journal of Fluid Mechanics}, 918, 2021.

\bibitem{Winn2024OperatingValves}
Aaron Winn and Eleni Katifori.
\newblock {Operating principles of peristaltic pumping through a dense array of valves}.
\newblock {\em Journal of Fluid Mechanics}, 989, July 2024.

\bibitem{Carr2021PeristalticTubes}
J.~Brennen Carr, John~H. Thomas, Jia Liu, and Jessica~K. Shang.
\newblock {Peristaltic pumping in thin non-axisymmetric annular tubes}.
\newblock {\em Journal of Fluid Mechanics}, 917, 2021.

\bibitem{Trevino2025LowSpace}
Avery Trevino, Thomas~R. Powers, Roberto Zenit, and Mauro Rodriguez.
\newblock {Low Reynolds number pumping near an elastic half space}.
\newblock {\em Physical Review Fluids}, 10(5), May 2025.

\bibitem{Burns_Parkes_1967}
J.~C. Burns and T.~Parkes.
\newblock Peristaltic motion.
\newblock {\em Journal of Fluid Mechanics}, 29(4):731–743, September 1967.

\bibitem{Taylor_1951}
Geoffrey Taylor.
\newblock Analysis of the swimming of microscopic organisms.
\newblock {\em Proceedings of the Royal Society of London. Series A. Mathematical and Physical Sciences}, 209(1099):447–461, November 1951.

\bibitem{Shapiro_Jaffrin_Weinberg_1969}
A.~H. Shapiro, M.~Y. Jaffrin, and S.~L. Weinberg.
\newblock {Peristaltic pumping with long wavelengths at low Reynolds number}.
\newblock {\em Journal of Fluid Mechanics}, 37(4):799–825, July 1969.

\bibitem{Jaffrin_Shapiro_1971}
M.~Y. Jaffrin and A.~H. Shapiro.
\newblock Peristaltic pumping.
\newblock {\em Annual Review of Fluid Mechanics}, 3(1):13–37, January 1971.

\bibitem{Li1993Non-SteadyTubes}
Meijing Li and James~G. Brasseur.
\newblock {Non-Steady Peristaltic Transport in Finite-Length Tubes}.
\newblock {\em Journal of Fluid Mechanics}, 248:129--151, 1993.

\bibitem{Pozrikidis_1987}
C.~Pozrikidis.
\newblock A study of peristaltic flow.
\newblock {\em Journal of Fluid Mechanics}, 180(1):515, July 1987.

\bibitem{Takabatake_Ayukawa_1982}
S.~Takabatake and K.~Ayukawa.
\newblock Numerical study of two-dimensional peristaltic flows.
\newblock {\em Journal of Fluid Mechanics}, 122(1):439, September 1982.

\bibitem{Takabatake_Ayukawa_Mori_1988}
S.~Takabatake, K.~Ayukawa, and A.~Mori.
\newblock Peristaltic pumping in circular cylindrical tubes: a numerical study of fluid transport and its efficiency.
\newblock {\em Journal of Fluid Mechanics}, 193(1):267, August 1988.

\bibitem{Tang1993NumericalBoundaries}
Dalin Tang and Samuel Rankin.
\newblock Numerical and asymptotic solutions for peristaltic motion of nonlinear viscous flows with elastic free boundaries.
\newblock {\em Siam Journal on Scientific Computing}, 14(6):1300--1319, November 1993.

\bibitem{Takagi_Balmforth_2011}
D.~Takagi and N.~J. Balmforth.
\newblock Peristaltic pumping of viscous fluid in an elastic tube.
\newblock {\em Journal of Fluid Mechanics}, 672:196–218, April 2011.

\bibitem{KouEtAl2015_ActiveMusculoMechanical}
Wenjun Kou, Amneet Pal~Singh Bhalla, Boyce~E. Griffith, John~E. Pandolfino, Peter~J. Kahrilas, and Neelesh~A. Patankar.
\newblock A fully resolved active musculo-mechanical model for esophageal transport.
\newblock {\em Journal of Computational Physics}, 298:446--465, 2015.

\bibitem{KouEtAl2015}
Wenjun Kou, John~E. Pandolfino, Peter~J. Kahrilas, and Neelesh~A. Patankar.
\newblock Simulation studies of circular muscle contraction, longitudinal muscle shortening, and their coordination in esophageal transport.
\newblock {\em American Journal of Physiology – Gastrointestinal and Liver Physiology}, 309(4):G238--G247, 2015.

\bibitem{AghilinejadEtAl2023}
Arian Aghilinejad, Bryson Rogers, Haojie Geng, and Niema~M. Pahlevan.
\newblock On the longitudinal wave pumping in fluid-filled compliant tubes.
\newblock {\em Physics of Fluids}, 35(9):091903, 2023.

\bibitem{Quillin1999KINEMATICTERRESTRIS}
Kim~J. Quillin.
\newblock Kinematic scaling of locomotion by hydrostatic animals: Ontogeny of peristaltic crawling by the earthworm lumbricus terrestris.
\newblock {\em The Journal of Experimental Biology}, 202:661--674, February 1999.

\bibitem{Elbaz_Gat_2014}
S.~B. Elbaz and A.~D. Gat.
\newblock Dynamics of viscous liquid within a closed elastic cylinder subject to external forces with application to soft robotics.
\newblock {\em Journal of Fluid Mechanics}, 758:221–237, November 2014.

\bibitem{Alim2013RandomIndividual}
Karen Alim, Gabriel Amselem, François Peaudecerf, Michael~P. Brenner, and Anne Pringle.
\newblock {Random network peristalsis in Physarum polycephalum organizes fluid flows across an individual}.
\newblock {\em Proceedings of the National Academy of Sciences of the United States of America}, 110(33):13306--13311, August 2013.

\bibitem{Aboelkassem2013SelectiveTransport}
Yasser Aboelkassem and Anne~E. Staples.
\newblock {Selective pumping in a network: Insect-style microscale flow transport}.
\newblock {\em Bioinspiration and Biomimetics}, 8(2), June 2013.

\bibitem{Southward1955ObservationsL.}
A.~J. Southward.
\newblock {Observations on the ciliary currents of the jelly-fish Aurelia aurita L.}
\newblock {\em Journal of the Marine Biological Association of the United Kingdom}, 34(2):201--216, 1955.

\bibitem{Tavakol2017ExtendedGeometry}
Behrouz Tavakol, Guillaume Froehlicher, Douglas~P. Holmes, and Howard~A. Stone.
\newblock {Extended lubrication theory: Improved estimates of flow in channels with variable geometry}.
\newblock {\em Proceedings of the Royal Society A: Mathematical, Physical and Engineering Sciences}, 473(2206), October 2017.

\bibitem{Walker2010ShapePumping}
Shawn~W. Walker and Michael~J. Shelley.
\newblock {Shape optimization of peristaltic pumping}.
\newblock {\em Journal of Computational Physics}, 229(4):1260--1291, February 2010.

\bibitem{Fung1993}
Y.~C. Fung.
\newblock {\em Biomechanics: Mechanical Properties of Living Tissues}.
\newblock Springer, New York, 2 edition, 1993.

\bibitem{Aris1989}
Rutherford Aris.
\newblock {\em Vectors, Tensors, and the Basic Equations of Fluid Mechanics}.
\newblock Dover Publications, New York, 1989.

\bibitem{1971RefluxVelocity}
A.~Shapiro and M.~Jaffrin.
\newblock {Reflux in Peristaltic Pumping: Is It Determined by the Eulerian or Lagrangian Mean Velocity?}
\newblock {\em Journal of Applied Mechanics}, pages 1060--1062, December 1971.

\bibitem{kurachi2000heart}
Y.~Kurachi, A.~Terzic, M.V. Cohen, and N.~Sperelakis.
\newblock {\em Heart Physiology and Pathophysiology}.
\newblock Academic Press, 2000.

\bibitem{Amselem2023ValvelessNumbers}
Gabriel Amselem, Christophe Clanet, and Michael Benzaquen.
\newblock {Valveless Pumping at Low Reynolds Numbers}.
\newblock {\em Physical Review Applied}, 19(2), February 2023.

\bibitem{Provost1994APumping}
Alden~M. Provost and W.~H. Schwarz.
\newblock {A Theoretical Study of Viscous Effects in Peristaltic Pumping}.
\newblock {\em Journal of Fluid Mechanics}, 279:177--195, 1994.

\bibitem{Canic2003EffectiveArteries}
Sunčica {\v{C}}ani{\'{c}} and Andro Mikeli{\'{c}}.
\newblock {Effective equations modeling the flow of a viscous incompressible fluid through a long elastic tube arising in the study of blood flow through small arteries}.
\newblock {\em SIAM Journal on Applied Dynamical Systems}, 2(3):431--463, September 2003.

\bibitem{Murata1976LaminarCurvature}
S.~Murata, Y.~Miyake, and T.~Inaba.
\newblock {Laminar flow in a curved pipe with varying curvature}.
\newblock {\em Journal of Fluid Mechanics}, 73(4):735--752, 1976.

\bibitem{Wang1981OnPipe}
C.~Y. Wang.
\newblock {On the low-Reynolds-number flow in a helical pipe}.
\newblock {\em Journal of Fluid Mechanics}, 108(7):185--194, 1981.

\bibitem{Donea2004Methods}
Jean Donea, Antonio Huerta, J.‐Ph. Ponthot, and A.~Rodr{\'{i}}guez‐Ferran.
\newblock {Arbitrary Lagrangian–Eulerian Methods}.
\newblock In {\em Encyclopedia of Computational Mechanics}. Wiley, August 2004.

\bibitem{Lauga2009TheMicroorganisms}
Eric Lauga and Thomas~R. Powers.
\newblock {The hydrodynamics of swimming microorganisms}.
\newblock {\em Reports on Progress in Physics}, 72(9), 2009.

\end{thebibliography}

\end{document}